\documentclass[pdflatex,sn-mathphys-num,iicol]{sn-jnl}

\usepackage{graphicx}%
\usepackage{multirow}%
\usepackage{amsmath,amssymb,amsfonts}%
\usepackage{amsthm}%
\usepackage{mathrsfs}%
\usepackage[title]{appendix}%
\usepackage{xcolor}%
\usepackage{textcomp}%
\usepackage{manyfoot}%
\usepackage{booktabs}%
\usepackage{array}%
\usepackage{rotating}%
\AtBeginDocument{\expandafter\def\expandafter\UrlBreaks\expandafter{\UrlBreaks\do-}}
\graphicspath{{./}{figures/}}



\theoremstyle{thmstyleone}%
\newtheorem{proposition}{Proposition}%

\begin{document}

\title[IPEK: Event-Criticality-Aware Evidential Trust Management]{IPEK: Event-Criticality-Aware Evidential Trust Management against Strategic Insider Attacks in Vehicular Ad Hoc Networks}

\author*[1]{\fnm{\.{I}pek} \sur{Abas{\i}kele\c{s}-Turgut}\,\orcid{https://orcid.org/0000-0002-5068-969X}}\email{ipek.abasikeles@iste.edu.tr}

\affil*[1]{\orgdiv{Computer Engineering Department}, \orgname{Iskenderun Technical University}, \orgaddress{\city{Hatay}, \postcode{31200}, \country{T\"{u}rkiye}}}

\abstract{%
Event-based trust management in vehicular ad hoc networks is vulnerable to strategic insider attackers who report trivial events correctly to build reputation, falsify reports on safety-critical events, and corrupt the trust feedback they provide about other vehicles. This paper proposes IPEK, an event-criticality-aware evidential trust management framework. IPEK combines an asymmetric local trust that scales rewards and penalties with event severity and location criticality, a reporter-credibility test that discounts feedback deviating systematically from the consensus of witnesses, and a Dempster--Shafer global trust that fuses feedback with Yager's rule, historical combination and risk accentuation. Closed-form properties quantify how quickly trust is lost and regained and show why Yager's rule keeps the global state revisable. IPEK is evaluated in simulation against a trust cascading-based dissemination scheme and the beta reputation system under eight attacks, with severity-, location-, on-off- and continuously triggered attackers falsifying their reports alone or also their feedback, at attacker ratios from 10\% to 40\%. Threshold-free separation measures are complemented by closed-loop revocation with frozen thresholds. With up to 25\% attackers, IPEK achieves a significantly higher F1-score than both baselines under all attacks, with false positive rates of 0.9\%--15.5\% against 7.8\%--49.7\% and 26.0\%--41.4\%, and it remains robust up to 40\% attackers when only reports are falsified. Under fully colluding attackers that also corrupt feedback, all methods break down beyond about one third of attackers; this envelope and a longer detection delay are reported as limitations.}


\maketitle

\section{Introduction}\label{sec:intro}
Real-time traffic information has become essential for modern transportation, with navigation platforms now serving billions of users worldwide through crowdsourced data collection \cite{abasikelecs2024recent,businessofapps2026,pichai2024alphabet,bradshaw2019googlemaps}. However, this reliance on user-generated reports introduces a critical reliability challenge, as the accuracy of information provided by potentially malicious participants cannot be inherently guaranteed. Vehicular Ad Hoc Networks (VANETs), as a key component of Intelligent Transportation Systems, address this challenge through inter-vehicle trust management mechanisms \cite{tcemd,notrino,duel,htemd,rteam,mdt,hdrs,marine,aatms,rsma,htms,misbehav}.

In recent years, VANET trust management has attracted significant interest in literature; behavioral trust, data-based trust, and hybrid approaches have been examined comprehensively \cite{abasikelecs2024recent}. Nevertheless, the vast majority of existing studies assume continuous message exchange between vehicles and compute trust values from the success/failure statistics of these messages. In real traffic environments, however, messages are generated only when specific events occur, resulting in a much sparser communication structure. Similarly, trust parameters derived from periodic messages such as beacons do not involve the direct witnessing of a specific event and lack contextual accuracy analysis. These limitations highlight event-based trust mechanisms as a more suitable framework for real traffic scenarios.

While event-based trust mechanisms are better suited to these requirements, existing models \cite{tcemd,notrino,duel,htemd,rteam,turgut2025effect} typically adopt a homogeneous treatment of events, thereby failing to account for the inherent differences in event severity and location criticality. This uniformity does not align with the heterogeneous nature of real traffic environments. On the other hand, although intelligent attacks have been modeled in the literature \cite{htms,notrino,htemd,mdt,hdrs,marine,aatms}, existing models are based on behavior patterns that change over time, such as trust-threshold-dependent on-off strategies. In these models, the attacker decides when to attack on the basis of time or of its own trust level; the importance of the event itself plays no role in the decision. An attacker that exploits event importance behaves differently: it reports low-importance events---such as potholes or light congestion---correctly to accumulate a high trust score, and then deliberately presents misleading information during a high-importance event---such as an accident in a school zone or a sudden road closure. Approaches that do not incorporate event type into trust calculation continue to evaluate this attacker as trustworthy owing to its high trust score. The threat is compounded when the same attackers also corrupt the trust feedback they provide about others, supporting one another and discrediting honest vehicles.

In this study, IPEK, an event-criticality-aware evidential trust management framework, is proposed. The following contributions are presented:

\textbf{1. Event- and location-criticality-aware local trust.} Extending the asymmetric reward-penalty approaches in the literature, a local trust mechanism that jointly evaluates event severity and location criticality is proposed. Rewards grow asymptotically and slowly, whereas penalties are amplified by a combined criticality factor and scaled by the reliability of the two observations, so that trust is gained slowly on trivial events and lost quickly on critical ones. Closed-form expressions quantify this asymmetry, and the benefit and the cost of criticality awareness are measured separately for attacks that follow criticality and for attacks that do not.

\textbf{2. Reporter credibility against feedback manipulation.} A deviation test assigns each reporter a credibility based on how systematically its feedback departs from the consensus of other witnesses, and the credibility discounts its evidence before fusion. The test detects colluding bad-mouthing and ballot stuffing from the pattern of the feedback rather than from its values.

\textbf{3. Evidential global trust with a measured choice of combination rule.} Feedback is converted into mass functions weighted by the reporter's global trust and credibility, fused sequentially with Yager's rule, combined with the history of each vehicle, and shifted toward risk when the current evidence is predominantly risky. The trade-off between Yager's and Dempster's rules---higher peak performance versus robustness to inversion---is explained analytically and quantified experimentally.

\textbf{4. A two-layer evaluation under strategic attacks.} IPEK is compared with TCEMD and beta reputation under eight attacks, combining severity-, location-, on-off- and continuously triggered attackers with dual-channel and report-only falsification, at attacker ratios from 10\% to 40\%. The evaluation combines threshold-free separation measures with closed-loop revocation at thresholds frozen in a reference configuration, and it locates the attacker share at which the methods break down.

The results show that, with up to 25\% attackers, IPEK achieves a significantly higher F1 than both baselines under all eight attacks, while keeping its false positive rate between 0.9\% and 15.5\%, compared with 7.8\%--49.7\% for beta reputation and 26.0\%--41.4\% for TCEMD. Against attackers that falsify only their reports, IPEK remains robust up to 40\% attackers. Under fully colluding attackers that also corrupt their feedback, all compared methods break down once the attackers make up about one third of the vehicles; this operating envelope, together with a longer detection delay than beta reputation and a weaker performance against on-off attacks at higher attacker ratios, is reported as a measured limitation.

The remainder of the paper is organized as follows. Section~\ref{sec:related} reviews the related literature. Section~\ref{sec:model} presents the system and threat models. Section~\ref{sec:ipek} describes the IPEK framework, and Section~\ref{sec:analysis} derives its analytical properties. Section~\ref{sec:method} details the evaluation methodology, and Section~\ref{sec:results} reports the results. Section~\ref{sec:discussion} discusses the findings and limitations, and Section~\ref{sec:conclusion} concludes the paper.
\section{Related Work}\label{sec:related}
VANET trust systems are classified from various perspectives in the literature: data-centric and entity-centric according to the evaluated target, and centralized and distributed according to the design architecture. These classifications and a general literature review are presented in detail in \cite{abasikelecs2024recent}. In this study, the evaluation is conducted in terms of the fundamental components of event-based trust systems: local trust calculation, global trust calculation and attack models. Table~\ref{tab:related} compares the design choices of recent studies across these three dimensions and their architectural type.

\begin{sidewaystable*}
\caption{Design choices and parameters of recent trust management schemes}\label{tab:related}
\footnotesize
\setlength{\tabcolsep}{2pt}
\begin{tabular}{p{1.6cm} p{3.0cm} p{1.9cm} p{2.2cm} p{2.2cm} p{2.8cm} p{1.9cm} p{1.8cm}}
\toprule
\textbf{Related Work} & \textbf{LT Parameters} & \textbf{GT Parameters} & \textbf{LT Alg.} & \textbf{GT Alg.} & \textbf{Attacks} & \textbf{System Type} & \textbf{Architecture} \\
\midrule
TCEMD \cite{tcemd} & DO & LT, Message Stats & No algorithm (binary value) & Weighted average & MITM, False Data & Event-based & Centralized \\
MDT \cite{mdt} & DO (comm. success, msg. validation), NR & LT, HT & No aggregation (per par. score) & Weighted average, MAD & Recommendation, Intelligent, Black Hole & Continuous msg-based & Centralized \\
HDRS \cite{hdrs} & DO (comm. success), NR, Role-based rules, GT & LT, HT, NR & Weighted average & Weighted average & Recommendation, Intelligent, False Data, Selfish & Continuous msg-based & Distributed \\
NOTRINO \cite{notrino} & Distance, Antenna height, Role-based rules & -- & Arithmetic operations & -- & Intelligent, MITM & Event-based & Distributed \\
DUEL \cite{duel} & NR, Time, Message rate, Modify rate, HT & -- & DST & -- & MITM, False Data & Event-based & Distributed \\
MARINE \cite{marine} & Role-based rules, Message Stats, Distance, NR & LT, HT & Arithmetic operations & Weighted average & Intelligent, MITM & Continuous msg-based & Centralized \\
AATMS \cite{aatms} & DO (comm. success) & LT, HT, Social Factors & Bayesian & Weighted average & Recommendation, Intelligent, Newcomer & Continuous msg-based & Centralized \\
HTEMD \cite{htemd} & DO, GT & LT, HT & Weighted average & Weighted average, MAD & Recommendation, Intelligent, Black Hole & Event-based & Centralized \\
RSMA \cite{rsma} & DO (comm. success) & LT, HT & No aggregation (per par. score) & Weighted average & False Data & Continuous msg-based & Centralized \\
HTMS \cite{htms} & DO (comm. success) & LT, HT & Arithmetic operations & DST & Recommendation, Intelligent & Continuous msg-based & Centralized \\
\cite{misbehav} & Behavior metrics, Reputation & LT & SVM & DST & False Data, Black Hole & Continuous msg-based & Centralized \\
\bottomrule
\end{tabular}

\smallskip
LT: Local Trust, GT: Global Trust, DO: Direct Observation, NR: Neighbor Recommendation, HT: Historical Trust, MAD: Median Absolute Deviation, DST: Dempster--Shafer Theory, SVM: Support Vector Machine, MITM: Man-in-the-Middle
\end{sidewaystable*}

\textbf{Attack evaluation.} As shown in Table~\ref{tab:related}, traditional attacks, including false data injection, recommendation and man-in-the-middle (MITM) attacks, have been extensively evaluated in the literature. In addition, intelligent attack models, which perform attacks during specific time periods based on trust value, have also been proposed. However, existing intelligent models focus primarily on temporal strategies (e.g., time-based on-off) and remain oblivious to the contextual importance of event types or location criticality. This has prevented the development of attack models that can behave strategically based on event or location severity.

\textbf{Local trust evaluation.} Parameters used for local trust calculation typically assume continuous message exchange between vehicles, deriving trust values from statistical distributions based on the success/failure rates of these messages. This assumption is incompatible with real-world communication structures; in event-based trust systems the messages underlying trust evaluation are typically sparse and generated only when specific events occur. Trust calculations based on the continuous-message assumption may misestimate system reliability by disregarding the low frequency of events. In some studies, additional trust parameters are computed through the verification of periodic messages such as beacons (e.g., MDT \cite{mdt}); however, in these approaches a specific event is not directly witnessed, and periodic messages are included in trust calculation without contextual accuracy analysis. Among event-based systems, RTEAM \cite{rteam}, NOTRINO \cite{notrino}, TCEMD \cite{tcemd}, HTEMD \cite{htemd} and DUEL \cite{duel} stand out. RTEAM focuses on the decision-making process of vehicles facing conflicting reports but does not present a holistic trust-system design and does not consider event diversity. NOTRINO lacks mechanisms for event verification and event-based vehicle action. TCEMD and HTEMD perform trust calculation based on message broadcasting; both use a single type of event message, and since TCEMD does not take past evaluations into account, it produces independent values in each round, which results in fluctuations in global trust values. In summary, existing models largely treat events homogeneously and do not reflect differences in event severity and location criticality, leaving a serious vulnerability against intelligent attacks that behave strategically based on event importance.

\textbf{Global trust evaluation.} In combining data from different sources, the use of the median absolute deviation (MAD) or Dempster--Shafer theory (DST) is observed for filtering out abnormal values. DST provides a strong mathematical framework for combining conflicting and uncertain evidence. Its ability to explicitly model uncertainty through the interval between belief and plausibility enables a more robust calculation of global trust. However, the use of DST in the schemes summarized in Table~\ref{tab:related} has remained basic. The classical Dempster combination rule~\cite{dempster1967,shafer1976} can produce counter-intuitive results under high conflict~\cite{zadeh1986}, and rules that redistribute conflict differently, such as Yager's rule~\cite{yager1987} and the proportional conflict redistribution rules~\cite{smarandache2005pcr}, offer alternatives whose behavior in vehicular trust management remains largely unexplored. Evidence theory has also been used to establish trust in individual pieces of data in vehicular networks~\cite{raya2008datacentric}; that approach focuses on the trustworthiness of each piece of data rather than on the long-term reputation of the reporters. Likewise, the explicit treatment of conflicting evidence, a special emphasis on risky evidence, and sequential combination ordered by source reliability are not part of the schemes in Table~\ref{tab:related}.

\textbf{Probabilistic reputation and unfair feedback.} Outside the vehicular domain, the beta reputation system~\cite{josang2002beta} aggregates binary feedback into the expected value of a beta distribution, and subjective logic~\cite{josang2016sl} generalizes this model to opinions with explicit uncertainty. These models are simple, well founded and widely used, and the beta reputation system is therefore adopted as a baseline in this study. Both are, however, vulnerable to unfair feedback: colluding reporters can inflate the reputation of accomplices (ballot stuffing) and deflate that of honest participants (bad-mouthing). Proposed defenses filter or discount reporters whose feedback deviates from that of the majority, either statistically~\cite{whitby2004filtering} or through clustering~\cite{dellarocas2000immunizing}. Such majority-based filters fail when colluders become the majority, a limitation that also applies to the reporter credibility of IPEK and is quantified in Section~\ref{subsec:break}. In the vehicular schemes of Table~\ref{tab:related}, recommendation attacks are evaluated, but the feedback channel is rarely examined together with event-triggered false reporting, i.e., with attackers that manipulate both channels only when the event is worth attacking.

In summary, existing approaches exhibit four key limitations: (i) attack models that ignore event and location context, (ii) local trust mechanisms that treat all events homogeneously, (iii) basic DST implementations that struggle with high-conflict scenarios, and (iv) feedback-channel attacks that are rarely evaluated together with event-triggered false reporting. IPEK addresses these limitations through the mechanisms detailed in Section~\ref{sec:ipek}.
\section{System and Threat Model}\label{sec:model}
This section presents the components and operation of the system, the underlying assumptions, the event and perception model, and the threat model against which IPEK is designed and evaluated.

\subsection{System Components and Operation}\label{subsec:components}
\begin{figure*}[!t]
\centering
\includegraphics[width=\textwidth]{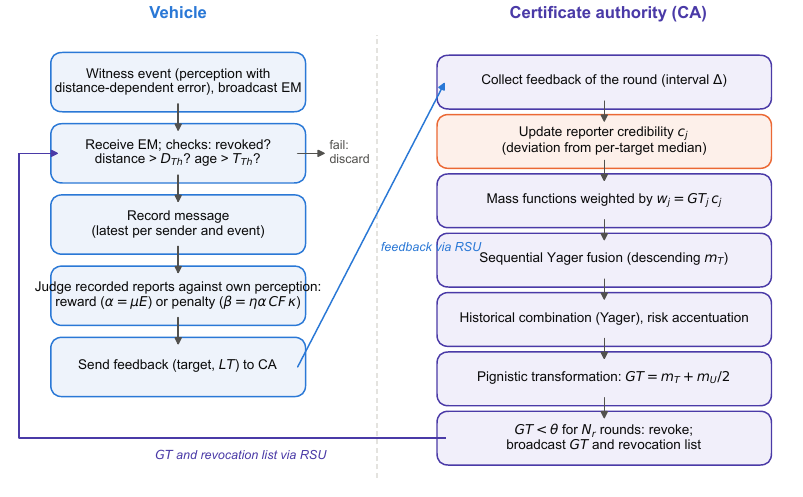}
\caption{End-to-end operational flow of the proposed system. On the vehicle side (left): witnessing an event, preliminary checks (revocation, distance and time thresholds), message recording, asymmetric local-trust computation and reporting of feedback to the CA. On the CA side (right): reporter credibility update, mass construction weighted by global trust and credibility, sequential Yager fusion, historical combination and risk accentuation, the pignistic transformation to a scalar global trust, revocation, and broadcasting of the global trust and revocation list. The two sides communicate through the RSU.}
\label{fig:flow}
\end{figure*}

The system consists of three main components: vehicles, road-side units (RSUs), and central authority (CA). RSUs are responsible for data transmission between vehicles and the CA. Since the modules contributing to trust calculation are the vehicles and the CA, the operation is defined over these two components: when a vehicle witnesses an event, it computes local trust and reports it to the CA through the RSUs; the CA periodically computes global trust and broadcasts it back to the vehicles through the RSUs. The end-to-end operation of the system is summarized in Fig.~\ref{fig:flow}.

\begin{figure}[!t]
\centering
\includegraphics[width=\columnwidth]{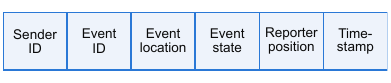}
\caption{The components of an event message (EM).}
\label{fig:em}
\end{figure}

Vehicles that witness an event, i.e., vehicles within its impact radius, alert their neighbors by broadcasting an event message (EM) that carries their perceived state of the event; the absence of an event is reported in the same way. A witness repeats this broadcast periodically (every 5\,s in this study) for as long as it remains within the impact radius. The EM structure and content are shown in Fig.~\ref{fig:em}. The message includes the sender's identity, the event identifier and location, the reported event state, the sender's position at the time of the report, and the time the message was sent. When a vehicle witnesses an event, it evaluates the vehicles that sent the EMs it has previously received and recorded, and calculates local trust for each of them. The calculated local trust values are transmitted to the CA through the RSU. As long as the vehicle continues to witness the event, it updates the local trust values as it receives new event messages and reports them to the CA.

Vehicles that receive an EM apply several preliminary checks before recording the message. First, whether the sending vehicle has been revoked is checked; this information is obtained from the revocation list broadcast by the CA after global trust computation. If the sender is not revoked, a distance check is performed: if the receiving vehicle is farther than an ignore threshold ($D_{Th}$) from the event, the message is disregarded. Similarly, if the message is older than a time threshold ($T_{Th}$), it is discarded. Messages passing these checks are recorded; for the same event from the same sender, only the most recent message is retained. In this study, the distance threshold $D_{Th}$ is set to 800\,m, so that only vehicles for which the event is locally relevant keep its reports, while the time threshold $T_{Th}$ is set dynamically to half of the event's active duration ($50 \cdot (1 + CF)$\,s, Section~\ref{subsec:events}), so that stale reports---which may no longer reflect the current state of the event---are discarded. The dynamic time threshold grants longer validity to reports about higher-criticality events, which remain active longer. The same age limit is applied when a witness evaluates recorded reports, so that a stale report is not penalized as false.

\subsection{Assumptions}\label{subsec:assumptions}
The operation of the proposed system rests on the following assumptions. Each vehicle is assumed to have a unique identity number and to use it in its communications. Since the aim of this work is to provide a solution against event-based internal intelligent attacks, external attacks are out of scope; certificate-based methods used in similar studies can be integrated into IPEK. The CA and RSUs are assumed to be trustworthy, free of resource constraints, and always accessible.

Vehicles are assumed to determine their own coordinates accurately via the Global Positioning System (GPS). However, vehicles are not assumed to perceive events perfectly. To reflect realistic sensing conditions, the probability that a vehicle misperceives the state of an event is modeled by a noise term that increases with the vehicle's distance from the event (Section~\ref{subsec:events}). This more realistic assumption creates an environment in which even honest vehicles may occasionally produce erroneous reports, thereby testing---under more demanding conditions---the trust mechanism's ability to distinguish deliberate attacks from innocent perception errors.

Message delivery within the broadcast range is assumed to be reliable: the evaluation focuses on the trust logic, and channel effects such as packet loss, contention and propagation delay are not modeled (Section~\ref{sec:method}).

\subsection{Event and Perception Model}\label{subsec:events}
Throughout the simulation, 40 traffic events occurring at predefined locations and times are modeled. To ensure spatial diversity, the network area is divided into four equal quadrants and events are distributed across them in turn; the exact location of each event is chosen uniformly at random within its quadrant and remains fixed throughout the simulation.

Each event is characterized by an \emph{event severity} ($S_E$) and a \emph{location criticality} ($S_L$). Events are grouped into three classes according to the levels of these two dimensions (Table~\ref{tab:events}): low-priority/non-critical events (Class~1), mixed events in which only one dimension is critical (Class~2), and events critical in both event and location (Class~3). The two dimensions are combined into a single coefficient that defines the combined criticality of an event, computed with the same formula used in the local-trust penalty (Section~\ref{subsec:local}): $CF = S_E + S_L - S_E \cdot S_L$.

\begin{figure*}[!t]
\centering
\includegraphics[width=\textwidth]{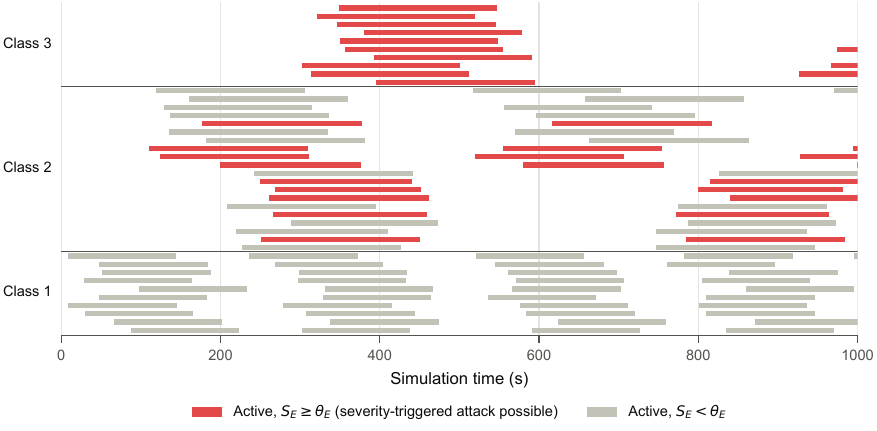}
\caption{Representative realization of the event activation timeline, grouped by criticality class and generated from the event model described in this section; the schedule is stochastic and seed-dependent, so the figure illustrates the model rather than a recorded run. Each bar denotes an active window; red windows mark events for which the severity-triggered attack condition holds ($S_E \geq \theta_E$), and gray windows mark active events on which such an attacker behaves honestly.}
\label{fig:timeline}
\end{figure*}

The combined criticality also determines how long an event remains active: the active duration is $100 \cdot (1 + CF)$ seconds, so that higher-criticality events remain active longer. Events are initially passive; after a random delay they become active, then passive again once their active duration elapses, and reactivate---at the same location and with the same severity---after a class-dependent random delay. Since the reactivation delay depends on the event class, low-criticality events recur frequently while high-criticality events recur rarely. This cyclic structure creates a demanding environment for trust models, testing long-term resilience by allowing an attacker to accumulate reputation during low-criticality periods and exploit it during high-criticality windows. A representative realization of the temporal distribution of event activations is shown in Fig.~\ref{fig:timeline}.

An event's impact radius is 250\,m, kept smaller than the vehicles' broadcast range, so that the set of witnesses that directly observe an event is a subset of the set of listeners that merely receive report messages. Vehicle perception is imperfect: the probability that a vehicle misperceives an event increases with its distance from the event and is modeled by~\eqref{eq:perc}:
\begin{equation}
p_{\text{err}}(d) = p_0 \cdot \frac{d}{R}
\label{eq:perc}
\end{equation}
where $d$ is the vehicle--event distance, $R$ the impact radius (250\,m), and $p_0$ the base error probability ($p_0 = 0.1$). The perception error is thus near zero at the event center and reaches $p_0$ at the edge of the impact radius. Honest vehicles report their perceived state (subject to this noise), whereas attackers falsify their reports under the conditions defined in Section~\ref{subsec:threat}.

\begin{table*}[!t]
\centering
\caption{Properties of the events modeled in the simulation}\label{tab:events}
\begin{tabular}{llp{3.2cm}p{4.2cm}}
\toprule
\textbf{Class} & \textbf{Event ID} & \textbf{$S_E$ / $S_L$} & \textbf{Description} \\
\midrule
Class 1 & 0--9 & 0.2 / 0.2 & Low-priority / non-critical event and location \\
Class 2 & 10--29 & one high [0.7--1.0], other low [0.1--0.6] & Only one dimension critical \\
Class 3 & 30--39 & 0.9 / 0.9 & High-priority / critical event and location \\
\bottomrule
\end{tabular}
\end{table*}

\subsection{Threat Model}\label{subsec:threat}
The adversary is an insider: a legitimate vehicle with a valid identity that follows the protocol in form but falsifies its content. At the start of a run, each vehicle independently becomes an attacker with probability $\pi_a$, the nominal attacker ratio. The realized share of attackers therefore varies from run to run; with 150 vehicles and $\pi_a = 0.25$, for example, it ranges from 30 to 44 attackers over the ten seeds used in the evaluation. As a worst case, attackers are assumed to know one another and to collude fully.

An attacker acts through up to two channels. On the \emph{report channel}, when its trigger condition holds for an active event that it witnesses, it broadcasts the event as absent, hiding an existing hazard; otherwise it reports its perceived state like an honest vehicle. On the \emph{feedback channel}, when the trigger condition holds for the event under evaluation, it replaces the local trust values it sends to the CA with extreme values: 0 for honest vehicles and 1 for accomplices. This is the strongest manipulation expressible in the feedback format of every compared method, so that all methods face attacks of identical strength. Outside the trigger condition, the attacker behaves honestly on both channels and thus accumulates reputation.

Four trigger conditions are considered. The \emph{severity} trigger attacks when the event severity reaches a threshold ($S_E \geq \theta_E$, $\theta_E = 0.6$) and represents the strategic attacker targeted by IPEK, which reports trivial events correctly and lies on important ones. The \emph{location} trigger attacks when the location criticality reaches a threshold ($S_L \geq \theta_L$, $\theta_L = 0.4$) and exploits the location dimension instead; with the event classes of Table~\ref{tab:events}, it also covers the Class~2 events of intermediate location criticality and therefore attacks a broader set of events. The \emph{on-off} trigger ignores event importance and attacks periodically, during the first half of every 40\,s period. The \emph{continuous} trigger attacks at every opportunity. The last two triggers do not follow event importance; they test whether the criticality awareness of IPEK remains useful, or becomes a burden, when the attack is unrelated to the variable that IPEK exploits.

Combining the four triggers with the two channel configurations yields the eight attacks of Table~\ref{tab:attacks}. In the \emph{dual-channel} attacks, the attacker falsifies both its reports and its feedback; in the \emph{report-only} attacks, its feedback is honest. Attacks that fabricate non-existent events, attackers that stay silent, Sybil identities and external attackers are outside the scope of this study.

\begin{table*}[!t]
\centering
\caption{Attacks used in the evaluation}
\label{tab:attacks}
\begin{tabular}{lll}
\toprule
\textbf{Trigger} & \textbf{Dual-channel (report + feedback)} & \textbf{Report-only} \\
\midrule
Severity, $S_E \geq 0.6$ & D-Sev & R-Sev \\
Location, $S_L \geq 0.4$ & D-Loc & R-Loc \\
On-off, 40\,s period, 50\% duty & D-OnOff & R-OnOff \\
Continuous & D-Cont & R-Cont \\
\bottomrule
\end{tabular}
\end{table*}
\section{The IPEK Framework}\label{sec:ipek}
IPEK consists of four components that operate on two sides of the system (Fig.~\ref{fig:flow}). On the vehicle side, a witness compares the reports it has received about an event with its own perception and updates the \emph{local trust} of each reporter with an asymmetric, criticality-aware rule (Section~\ref{subsec:local}). The updated values are sent to the CA as feedback. On the CA side, the feedback of each round is first used to update the \emph{credibility} of every reporter (Section~\ref{subsec:cred}); the feedback is then converted into mass functions weighted by the reporter's global trust and credibility and fused into an \emph{evidential global trust} (Section~\ref{subsec:global}); finally, vehicles whose global trust remains below a threshold are \emph{revoked} (Section~\ref{subsec:revocation}). The parameter values used in this study are listed in Table~\ref{tab:ipek_params}.

\subsection{Local Trust}\label{subsec:local}
Since local trust is computed at the vehicle level, a lightweight update rule is adopted to keep computational complexity low. A vehicle $V_j$ acts as a judge only for events it witnesses: when $V_j$ is within the impact radius of an event and holds its own perception of the event state, it compares every recorded report about that event with this perception. Reports older than the time threshold $T_{Th}$ (Section~\ref{subsec:components}) are not evaluated, so that a stale report is not penalized as false. The judge has no access to the ground truth; a report is considered correct when it agrees with the judge's own, possibly erroneous, perception. Each evaluated report updates the local trust $LT_{ji} \in [0,1]$ that $V_j$ holds for the reporter $V_i$; a vehicle that has not been evaluated before starts from the neutral value $LT_{ji} = 0.5$.

The event severity ($S_E$) and location criticality ($S_L$) of the event are used in both the reward and the penalty. The mechanism is asymmetric: gaining trust is markedly harder than losing it. This principle, embedded in the reward and penalty structures defined below, forms the first line of defense against attackers that build reputation on low-importance events.

\textbf{Reward.} When the report of $V_i$ agrees with the perception of $V_j$, a reward is applied. The importance of the event for the reward is expressed by a weighted average of the two criticality dimensions \eqref{eq:E}, which allows their relative contributions to be controlled through the weights $\omega_E$ and $\omega_L$:
\begin{equation}
E = \omega_E S_E + \omega_L S_L, \qquad \omega_E + \omega_L = 1
\label{eq:E}
\end{equation}
The reward rate is $\alpha = \mu E$, and local trust grows asymptotically toward its upper bound \eqref{eq:reward}:
\begin{equation}
LT_{ji} \leftarrow LT_{ji} + (1 - LT_{ji})\, \alpha
\label{eq:reward}
\end{equation}
Here, $(1 - LT_{ji})$ is the remaining distance between the current trust value and the maximum, creating an asymptotic growth model in which the gain decreases as trust increases. A low-trust vehicle obtains a relatively high gain for a correct report, while an already high-trust vehicle obtains a lower gain for the same behavior. This design naturally prevents the trust value from exceeding the upper bound and makes reaching high trust levels progressively more difficult. The balance coefficient $\mu$ determines how much of the remaining distance can be gained with a single correct report; a low $\mu$ makes trust gain slow and gradual, increasing the system's resistance to manipulation. Since $E$ is larger for critical events, a correct report on a critical event earns more trust than a correct report on a trivial one.

\textbf{Penalty.} When the report of $V_i$ contradicts the perception of $V_j$, local trust is decreased in proportion to the importance of the event and location. To this end, a combined factor ($CF$), based on the union principle of independent events in probability theory, is defined. $S_E$ and $S_L$ denote the probability of the corresponding dimension being critical; accordingly, $(1-S_E)$ and $(1-S_L)$ are the probabilities of not being critical. The probability that neither dimension is critical is given, under the independence assumption, by $(1-S_E)(1-S_L)$, and the combined factor is its complement \eqref{eq:CF}, representing the situation in which at least one dimension is critical:
\begin{equation}
CF = 1 - (1 - S_{E})(1 - S_{L}) = S_{E} + S_{L} - S_{E} S_{L}
\label{eq:CF}
\end{equation}
This probabilistic union ensures that a high severity in either the event type or the location is sufficient to trigger a significant penalty, reflecting a safety-critical perspective; the interaction term $S_E S_L$ prevents double counting and keeps the result within $[0,1]$. The penalty rate is tied to the reward rate of the same event through a penalty ratio $\eta > 1$ \eqref{eq:beta}, and local trust decays multiplicatively \eqref{eq:penalty}:
\begin{equation}
\beta = \min\left(1,\; \eta\, \alpha\, CF\, \kappa_{ji}\right)
\label{eq:beta}
\end{equation}
\begin{equation}
LT_{ji} \leftarrow (1 - \beta)\, LT_{ji}
\label{eq:penalty}
\end{equation}
Decreasing the current value rather than resetting it to a fixed level ensures that repeated incorrect reports are penalized cumulatively; sustained malicious behavior thus drives a vehicle's trust steadily toward zero. Since the reward and the penalty use the same importance information but the penalty is amplified by $\eta\,CF$, a single false report on a critical event erases the gain of many correct reports on trivial events. With the values in Table~\ref{tab:ipek_params}, $\eta\, \alpha\, CF \leq \eta \mu = 0.75$, so the clipping in \eqref{eq:beta} is never active.

\textbf{Observation reliability.} Since honest vehicles may misperceive an event (Section~\ref{subsec:events}), a disagreement between two vehicles is not necessarily evidence of deliberate falsification. The penalty is therefore scaled by an observation-reliability factor \eqref{eq:kappa}:
\begin{equation}
\kappa_{ji} = \max\left(0,\; (1 - 2p_J)(1 - 2p_R)\right)
\label{eq:kappa}
\end{equation}
where $p_J = p_{\text{err}}(d_J)$ is the perception-error probability of the judge at the moment of its perception, $p_R = p_{\text{err}}(\min(d_R, R))$ is the corresponding probability for the reporter computed from the position carried in the report, and $p_{\text{err}}$ is given by \eqref{eq:perc}. The factor equals one when both vehicles observe the event from its center and decreases as either observation becomes less reliable. It is applied to the penalty only.

In the severity-agnostic variant used in the evaluation (Section~\ref{subsec:baselines}), $E$ and $CF$ are replaced by the constants 0.55 and 0.80, respectively.

After the evaluation, $V_j$ sends the updated local trust values to the CA as a single feedback message containing (target, local trust) pairs. Feedback is generated only by witnesses and only when they hold reports to evaluate; IPEK therefore introduces no message types beyond the event reports and the feedback used by the compared methods.

\subsection{Reporter Credibility}\label{subsec:cred}
The global trust of a reporter (Section~\ref{subsec:global}) reflects the correctness of its \emph{event reports}, not the honesty of its \emph{feedback}. An attacker that reports events correctly most of the time keeps a high global trust and can use it to inject corrupted feedback, praising its accomplices and discrediting honest vehicles. The value of a single feedback item does not reveal the manipulation: a low local trust reported by an honest witness that has just caught a false report on a critical event is indistinguishable, as a value, from a low value fabricated by an attacker. What distinguishes the two is the \emph{pattern} across targets. An honest witness deviates from the opinion of other witnesses only for the few targets it has caught, whereas a colluding attacker deviates systematically, rating honest vehicles low and its accomplices high irrespective of the evidence. IPEK captures this pattern with a deviation test that assigns each reporter $V_j$ a credibility $c_j \in [0,1]$, initialized to $c_j = 1$.

In each round, before the mass functions are constructed, the CA processes the round's feedback in three steps; revoked reporters and targets are excluded.
\begin{enumerate}
\item For each target $V_i$, the consensus $\tilde{\ell}_i$ is the unweighted median of the local trust values reported about $V_i$ in the round. The consensus is defined only when at least three reporters rated $V_i$.
\item For each reporter $V_j$, let $H_j$ be the set of targets rated by $V_j$ for which a consensus is defined. If $|H_j| \geq 3$, the mean absolute deviation of the reporter from the consensus is
\begin{equation}
d_j = \frac{1}{|H_j|} \sum_{i \in H_j} \left| LT_{ji} - \tilde{\ell}_i \right|.
\label{eq:dev}
\end{equation}
\item The credibility is updated by exponential smoothing \eqref{eq:cred}; reporters with $|H_j| < 3$ keep their previous credibility:
\begin{equation}
c_j \leftarrow (1 - \gamma)\, c_j + \gamma \max\left(0,\; 1 - \frac{d_j}{\delta_0}\right)
\label{eq:cred}
\end{equation}
\end{enumerate}
Here $\delta_0$ is the deviation at which the instantaneous credibility reaches zero and $\gamma$ is the smoothing factor. The updated credibility is used in the same round to discount the reporter's feedback (Section~\ref{subsec:global}). The approach follows the majority-based filtering of unfair ratings in reputation systems~\cite{whitby2004filtering,dellarocas2000immunizing}. The median is used instead of the mean since it is not shifted by a minority of extreme values; the test is therefore effective as long as colluding attackers are a minority among the reporters of a target, and it loses its reference when they become the majority. The parameters $\gamma$ and $\delta_0$ were fixed before the evaluation and were not tuned (Section~\ref{subsec:sens}).

\subsection{Evidential Global Trust}\label{subsec:global}
Global trust is computed on the centralized CA, which is assumed to be free of resource constraints. Dempster-Shafer Theory (DST) is adopted to manage conflicting reports from different vehicles, as it offers the capacity to explicitly model uncertainty beyond classical probability theory. When sufficient evidence about a vehicle is unavailable, the system can represent this as ``uncertain'' rather than forcibly classifying the vehicle as ``trusted'' or ``risky''---a property particularly suited to VANET environments, where multiple vehicles may report different observations about the same target.

Within DST, a three-component mass function is defined for each vehicle over the frame of discernment consisting of hypotheses T (trusted) and R (risky): trusted ($m_T$), risky ($m_R$) and uncertain ($m_U$), satisfying the normality condition $m_{T} + m_{R} + m_{U} = 1$. Global trust values are stored in mass-function form so that uncertainty information is preserved and can be used in subsequent fusion. Vehicles newly joining the network are assigned complete uncertainty ($m_T = m_R = 0$, $m_U = 1$), ensuring an unbiased evaluation process in which trustworthiness is determined solely by observed behavior. Whenever a scalar trust value is required, the mass function is converted to a single value in $[0,1]$ via the pignistic transformation.

The CA processes the feedback received during each update interval of length $\Delta$ (one \emph{round}); for each reporter--target pair, the most recent value in the round is used. When the local trust value $LT_{ji}$ of a reporting vehicle $V_j$ regarding a target $V_i$ is converted into a mass function, the reporter's weight $w_j = GT_j \cdot c_j$ is used as a discounting factor, where $GT_j$ is the reporter's current global trust and $c_j$ its credibility (Section~\ref{subsec:cred}). Consequently, feedback from low-trust or low-credibility reporters carries high uncertainty and has limited impact on the final decision. The components are computed by \eqref{eq:mT}--\eqref{eq:mU}:
\begin{align}
m_{T} &= w_{j} \cdot LT_{ji} \label{eq:mT} \\
m_{R} &= w_{j} \cdot (1 - LT_{ji}) \label{eq:mR} \\
m_{U} &= 1 - w_{j} \label{eq:mU}
\end{align}

Yager's rule is selected for fusing mass functions from multiple sources. Whereas the classical Dempster rule eliminates conflicting evidence through normalization, Yager's rule transfers conflict to the uncertainty set. Given that attackers may deliberately generate conflicting reports, this choice keeps the fused state revisable rather than committing it to a definite outcome on the basis of contested evidence. The combined trusted and risky masses are given by \eqref{eq:yT} and \eqref{eq:yR}, and the conflict factor $K$ by \eqref{eq:K}:
\begin{align}
m_{T}^{comb} &= m_{T}^{1} m_{T}^{2} + m_{T}^{1} m_{U}^{2} + m_{U}^{1} m_{T}^{2} \label{eq:yT} \\
m_{R}^{comb} &= m_{R}^{1} m_{R}^{2} + m_{R}^{1} m_{U}^{2} + m_{U}^{1} m_{R}^{2} \label{eq:yR} \\
K &= m_{T}^{1} m_{R}^{2} + m_{R}^{1} m_{T}^{2} \label{eq:K}
\end{align}

Under Yager's rule, this conflict is added to the uncertainty component rather than forcing a definitive decision \eqref{eq:yU}; the system thus awaits further evidence in the presence of conflicting reports:
\begin{equation}
m_{U}^{comb} = m_{U}^{1} m_{U}^{2} + K
\label{eq:yU}
\end{equation}

When multiple reports about the same target are available, they are sorted in descending order of the trusted mass $m_T$ of the reporters' global mass functions and fused pairwise and sequentially. This sequential fusion prevents low-trust nodes from polluting the baseline established by highly reliable sources; since Yager's rule is not associative, the order is part of the definition. With reporters ordered as $m_{T,(1)} \geq m_{T,(2)} \geq \cdots \geq m_{T,(n)}$ and $M^{(1)} = m^{(1)}$, the cumulative fusion proceeds by \eqref{eq:seq}:
\begin{equation}
M^{(k)} = \text{Yager}\left( M^{(k-1)}, m^{(k)} \right), \quad k = 2, \ldots, n
\label{eq:seq}
\end{equation}

If a previously calculated global trust value is available for a vehicle, the previous mass function ($M^{old}$) is combined with the current mass function ($M^{curr} = M^{(n)}$) using Yager's rule, as shown in \eqref{eq:hist}. No discounting is applied at this stage; discounting occurs only at the reporter level, when constructing mass functions from feedback via \eqref{eq:mT}--\eqref{eq:mU}. This prevents historical accumulation from being additionally and unintentionally penalized:
\begin{equation}
M^{new} = \text{Yager}\left( M^{old}, M^{curr} \right)
\label{eq:hist}
\end{equation}

After historical combination, an asymmetric risk accentuation mechanism is applied to enable the early detection of potential threats. When the risky component of the current round's fused mass $m_R^{curr}$ exceeds a threshold $\tau$, the result is shifted toward risk. The shift amount is obtained by scaling the excess over the threshold by a coefficient $\rho_{\text{risk}}$ \eqref{eq:boost}:
\begin{equation}
b = (m_{R}^{curr} - \tau) \cdot \rho_{\text{risk}}
\label{eq:boost}
\end{equation}

This shift is drawn primarily from the uncertainty component, since uncertainty represents a state in which definitive evidence has not yet formed and is therefore the preferred component to reduce first when risky evidence arrives. The amount taken from uncertainty $\Delta_U$ is the minimum of the current uncertainty and the required shift \eqref{eq:dU}; the risky component is increased and the uncertainty decreased by this amount \eqref{eq:dU2}:
\begin{equation}
\Delta_{U} = \min\left( m_{U}^{new}, b \right)
\label{eq:dU}
\end{equation}
\begin{equation}
m_{R}^{new} \leftarrow m_{R}^{new} + \Delta_{U}, \qquad
m_{U}^{new} \leftarrow m_{U}^{new} - \Delta_{U}
\label{eq:dU2}
\end{equation}

When the uncertainty component is insufficient to satisfy the required amount, the remainder is drawn from the trusted component, up to at most half of its value. Since the trusted component represents the accumulation of past positive behavior, this 50\% limit acts as a ``trust inertia'' that prevents a single negative report---or a potentially coordinated bad-mouthing attack---from instantly revoking a long-term honest participant \eqref{eq:dT}--\eqref{eq:dT2}:
\begin{equation}
\Delta_{T} = \min\left( 0.5\, m_{T}^{new}, b - \Delta_{U} \right)
\label{eq:dT}
\end{equation}
\begin{equation}
m_{R}^{new} \leftarrow m_{R}^{new} + \Delta_{T}, \qquad
m_{T}^{new} \leftarrow m_{T}^{new} - \Delta_{T}
\label{eq:dT2}
\end{equation}

Finally, the scalar global trust value is obtained via the pignistic transformation \eqref{eq:pign}:
\begin{equation}
GT_{j} = m_{T} + \frac{m_{U}}{2}
\label{eq:pign}
\end{equation}

The values of $\tau$, $\rho_{\text{risk}}$ and the trust inertia used in this study are listed in Table~\ref{tab:ipek_params}.

\subsection{Revocation}\label{subsec:revocation}
At the end of each round, the CA compares the global trust of every non-revoked vehicle with a detection threshold $\theta$. A vehicle is revoked when its global trust remains below $\theta$ for $N_r$ consecutive rounds; a round in which the global trust is at or above $\theta$ resets the count. Requiring several consecutive rounds prevents a single burst of negative feedback from revoking an honest vehicle, at the cost of a later reaction. Once revoked, a vehicle is placed on the revocation list: its event reports are discarded by receivers, its feedback is discarded by the CA, and feedback about it is no longer processed. The same revocation rule, with a method-specific threshold, is applied to all compared methods (Section~\ref{sec:method}).

\subsection{Computational and Communication Cost}\label{subsec:cost}
On the vehicle side, a witness performs a constant number of arithmetic operations per evaluated report. On the CA side, let $F$ denote the number of feedback items received in a round and $n_i$ the number of reporters of target $V_i$. The credibility step requires one median per target and one pass over the feedback, i.e., $O(\sum_i n_i \log n_i) \subseteq O(F \log F)$ operations; the sequential fusion requires sorting the reporters of each target and $n_i - 1$ constant-time combinations, which has the same order. The CA thus processes a round in $O(F \log F)$ time and stores one mass function and one credibility value per vehicle. In terms of communication, IPEK uses the same event reports and the same feedback messages as the compared methods; the only difference lies in the payload, which carries a continuous local trust value instead of a binary one.

\begin{table*}[!t]
\centering
\caption{IPEK parameters}
\label{tab:ipek_params}
\begin{tabular}{llc}
\toprule
\textbf{Symbol} & \textbf{Description} & \textbf{Value} \\
\midrule
\multicolumn{3}{l}{\textit{Local trust}} \\
$LT^{(0)}$ & Initial local trust & 0.5 \\
$\omega_E,\ \omega_L$ & Weights of event severity and location criticality in $E$ & 0.6, 0.4 \\
$\mu$ & Balance coefficient (reward rate scale) & 0.15 \\
$\eta$ & Penalty ratio & 5 \\
\midrule
\multicolumn{3}{l}{\textit{Reporter credibility}} \\
$c_j^{(0)}$ & Initial credibility & 1 \\
$\gamma$ & Smoothing factor & 0.5 \\
$\delta_0$ & Deviation at zero credibility & 0.5 \\
-- & Minimum reporters per target / targets per reporter & 3 / 3 \\
\midrule
\multicolumn{3}{l}{\textit{Global trust and revocation}} \\
$\tau$ & Risk-accentuation threshold & 0.5 \\
$\rho_{\text{risk}}$ & Risk-accentuation coefficient & 0.5 \\
-- & Trust inertia & 0.5 \\
$\theta$ & Detection threshold & 0.39 \\
\bottomrule
\end{tabular}
\end{table*}
\section{Analytical Properties}\label{sec:analysis}
This section derives properties of IPEK that follow directly from its update rules. They quantify how quickly a misbehaving vehicle loses trust, how slowly it regains it, what level of trust an honest but imperfect witness converges to, how credibility and fusion bound the influence of a reporter, and how the choice of combination rule affects the recoverability of the global state. Throughout, $\alpha = \mu E$ and $\beta = \eta \alpha\, CF\, \kappa$ denote the reward and penalty rates of an event (Section~\ref{subsec:local}); since $\eta\mu \leq 1$ in this study, $\beta < 1$ always holds.

\subsection{Local Trust Dynamics}

\begin{proposition}[Loss and recovery]\label{prop:lossrec}
Consider repeated evaluations of the same reporter on events of the same class. After $n$ consecutive false reports, the local trust is $LT_n = (1-\beta)^n LT_0$; after $n$ consecutive correct reports, it is $LT_n = 1 - (1 - LT_0)(1 - \alpha)^n$. Consequently, the number of false reports needed to bring the local trust from $LT_0$ below a level $\ell < LT_0$ is
\begin{equation}
n_{\text{loss}} = \left\lceil \frac{\ln (\ell / LT_0)}{\ln (1 - \beta)} \right\rceil,
\label{eq:nloss}
\end{equation}
and the number of correct reports needed to raise it from $\ell$ back to $LT_0$ is
\begin{equation}
n_{\text{rec}} = \left\lceil \frac{\ln \left( (1 - LT_0)/(1 - \ell) \right)}{\ln (1 - \alpha)} \right\rceil.
\label{eq:nrec}
\end{equation}
\end{proposition}
\begin{proof}
Both updates \eqref{eq:reward} and \eqref{eq:penalty} are affine maps with constant coefficients when the event class is fixed. Iterating $LT \leftarrow (1-\beta) LT$ gives the first expression, and iterating $1 - LT \leftarrow (1-\alpha)(1 - LT)$ gives the second. Solving $LT_n < \ell$ and $LT_n \geq LT_0$ for $n$, respectively, yields \eqref{eq:nloss} and \eqref{eq:nrec}.
\end{proof}

With the parameters of Table~\ref{tab:ipek_params} and $\kappa = 1$, a Class~3 event ($S_E = S_L = 0.9$) gives $\alpha = 0.135$ and $\beta = 0.668$: two false reports bring the local trust from the neutral value 0.5 below 0.1, whereas five correct reports are needed to return from 0.1 to 0.5. For a Class~1 event ($S_E = S_L = 0.2$), $\alpha = 0.030$ and $\beta = 0.054$, so that 29 false reports are needed for the same loss and 20 correct reports for the recovery. The ratio $\beta/\alpha = \eta\, CF\, \kappa$ ranges from 1.8 for Class~1 to 4.95 for Class~3. The penalty is thus mild for trivial events, on which a disagreement is more likely to be an innocent perception error and of little consequence, and severe for critical ones.

\begin{proposition}[Cross-class compensation]\label{prop:comp}
Let a reporter with local trust $LT_0$ lie once on an event with penalty rate $\beta_h$ and then report correctly on events with reward rate $\alpha_\ell$. The number of correct reports needed to restore $LT_0$ is
\begin{equation}
n_{\text{comp}} = \left\lceil \frac{\ln \left( (1 - LT_0)/(1 - (1-\beta_h) LT_0) \right)}{\ln (1 - \alpha_\ell)} \right\rceil.
\label{eq:ncomp}
\end{equation}
\end{proposition}
\begin{proof}
After the false report, $LT_1 = (1-\beta_h) LT_0$. Applying Proposition~\ref{prop:lossrec} with $\ell = LT_1$ and reward rate $\alpha_\ell$ gives \eqref{eq:ncomp}.
\end{proof}

Proposition~\ref{prop:comp} captures the strategic attacker that IPEK targets. A vehicle with local trust 0.8 that lies once on a Class~3 event drops to 0.265; regaining 0.8 requires 43 correct reports on Class~1 events, or nine on Class~3 events. Since the attacker of interest lies precisely on the critical events, it can only rebuild its reputation slowly on trivial ones, and each further lie on a critical event wipes out the accumulated gain. A severity-agnostic rule with constant rates offers no such leverage: a lie on a critical event costs the same as a lie on a trivial one.

\begin{proposition}[Expected-drift equilibrium]\label{prop:eq}
Suppose that each evaluation of a reporter on events of a given class is judged correct with probability $q$, independently of the past. The expected change of the local trust vanishes at
\begin{equation}
LT^{*} = \frac{q\,\alpha}{q\,\alpha + (1-q)\,\beta} = \frac{q}{q + (1-q)\,\eta\, CF\, \kappa},
\label{eq:eq}
\end{equation}
and the expected change is positive below $LT^{*}$ and negative above it.
\end{proposition}
\begin{proof}
From \eqref{eq:reward} and \eqref{eq:penalty}, $\mathbb{E}[\Delta LT] = q\,\alpha\,(1 - LT) - (1-q)\,\beta\, LT$, which is linear and decreasing in $LT$ and vanishes at $LT^{*}$.
\end{proof}

$LT^{*}$ is the point toward which the local trust drifts on average; it is not the mean of the stochastic process, which also depends on the order of the outcomes. It nevertheless shows the separation that the rule produces. An honest reporter is judged incorrect only when exactly one of the two vehicles misperceives the event, which happens with a probability $\epsilon = 1 - q$ of at most $2p_0 = 0.2$ and much smaller near the event center. For $\epsilon = 0.05$ and $\kappa = 1$, $LT^{*}$ is 0.79 for Class~3 events and 0.91 for Class~1 events; the observation-reliability factor ($\kappa < 1$) raises these values further, since the disagreements of honest vehicles concentrate near the edge of the impact radius where $\kappa$ is small. An attacker that lies whenever it is triggered has $q \approx 0$ on the triggered events and is driven toward zero on them.

\subsection{Influence of a Reporter}

\begin{proposition}[Single-source influence]\label{prop:single}
For a single feedback item with local trust $\ell$ and weight $w = GT_j c_j$, the pignistic value $m_T + m_U/2$ of the mass function \eqref{eq:mT}--\eqref{eq:mU} is
\begin{equation}
m_T + \frac{m_U}{2} = \frac{1}{2} + w \left( \ell - \frac{1}{2} \right).
\label{eq:single}
\end{equation}
\end{proposition}
\begin{proof}
By \eqref{eq:mT}--\eqref{eq:mU}, $m_T + m_U/2 = w\ell + (1 - w)/2$.
\end{proof}

A feedback item can move the trust of its target away from the neutral value by at most $w/2$. A reporter whose credibility has dropped to zero is therefore reduced to complete ignorance, regardless of its global trust, and the influence of the extreme values 0 and 1 used by colluding attackers is bounded by the weight of the reporter that sends them.

\begin{proposition}[Credibility bounds]\label{prop:cred}
If a reporter deviates from the consensus by $d_j \geq \delta_0$ in each of $n$ consecutive updates, then $c_j \leq (1-\gamma)^n c_j^{(0)}$. If a reporter deviates by at most $\bar{d} < \delta_0$ in every update, then $c_j \geq \min\left( c_j^{(0)},\, 1 - \bar{d}/\delta_0 \right)$ at all times.
\end{proposition}
\begin{proof}
By \eqref{eq:cred}, $c_j$ is a convex combination of its previous value and the instantaneous credibility $\max(0, 1 - d_j/\delta_0)$. In the first case the instantaneous credibility is zero and $c_j$ is multiplied by $1-\gamma$ at each update. In the second case the instantaneous credibility is at least $1 - \bar{d}/\delta_0$, and a convex combination of two values not smaller than a bound is not smaller than that bound.
\end{proof}

With $\gamma = 0.5$, a colluding attacker that consistently deviates from the consensus loses half of its remaining credibility in every round in which it is tested, whereas an honest witness whose average deviation is $\bar{d} = 0.1$ keeps a credibility of at least 0.8. The bound relies on the consensus being formed by honest reporters; when attackers become the majority among the reporters of the targets, the roles reverse (Section~\ref{subsec:break}).

\begin{proposition}[Bounded risk accentuation]\label{prop:risk}
In a single round, risk accentuation moves at most $(1 - \tau)\rho_{\text{risk}}$ of mass toward the risky hypothesis and removes at most half of the trusted mass.
\end{proposition}
\begin{proof}
Since $m_R^{curr} \leq 1$, \eqref{eq:boost} gives $b \leq (1-\tau)\rho_{\text{risk}}$, and $\Delta_U + \Delta_T \leq b$. By \eqref{eq:dT}, $\Delta_T \leq 0.5\, m_T^{new}$.
\end{proof}
With $\tau = \rho_{\text{risk}} = 0.5$, a round can shift at most 0.25 of mass toward risk, so that a single burst of negative feedback cannot by itself drive a vehicle with a substantial trusted mass to zero.

\subsection{Recoverability of the Global State}

\begin{proposition}[Dogmatic states]\label{prop:dogma}
Let the historical mass function be fully committed to the risky hypothesis, $M^{old} = (0, 1, 0)$, and let the current mass function be $M^{curr} = (t, r, u)$ with $t < 1$. Under Dempster's rule, $M^{new} = (0, 1, 0)$. Under Yager's rule, $M^{new} = (0,\, 1 - t,\, t)$.
\end{proposition}
\begin{proof}
With $M^{old} = (0,1,0)$, the unnormalized products are $m_T = 0$, $m_R = r + u = 1 - t$ and the conflict is $K = t$, with no product left on the uncertainty. Dempster's rule divides by $1 - K = 1 - t$ and returns $(0, 1, 0)$. Yager's rule assigns the conflict to uncertainty and returns $(0, 1 - t, t)$.
\end{proof}

Under Dempster's rule, a fully committed state is absorbing: no amount of positive evidence can change it. Near-committed states behave similarly, since normalization amplifies whichever hypothesis leads, so that beliefs sharpen quickly and recover slowly. Under Yager's rule, conflicting evidence reopens the uncertainty, from which the trusted mass can grow again in the next combination; the state remains revisable. This difference explains the trade-off observed in Section~\ref{subsec:rule}: Dempster's rule separates faster while honest evidence dominates, but locks in once the attackers prevail.

\subsection{Revocation Delay}
Since revocation requires $N_r$ consecutive rounds below the threshold and the global trust is updated once per round, a vehicle cannot be revoked earlier than $(N_r - 1)\Delta$ after the first round of the uninterrupted sequence of below-threshold rounds that leads to its revocation; with $N_r = 3$ and $\Delta = 50$\,s, this is 100\,s. The same bound applies to every method that uses this revocation rule, and $N_r$ directly trades reaction time against robustness to transient drops (Section~\ref{subsec:sens}).
\section{Evaluation Methodology}\label{sec:method}
This section presents the experimental environment and methodology used to evaluate IPEK and the compared methods. The evaluation is carried out in an integrated environment in which the discrete-event network simulator OMNeT++~\cite{varga2008omnet} is combined, through the TraCI interface of the Veins framework~\cite{sommer2011veins}, with SUMO~\cite{lopez2018sumo} for realistic vehicular mobility. Messages are delivered directly to all receivers within the broadcast range, without a physical- or medium-access-layer model, so that the comparison isolates the trust logic from channel effects; consequently, no claim is made about latency, packet loss or channel load.

\subsection{Simulation Environment}\label{subsec:simenv}
The road network is designed as a 5 $\times$ 5 Manhattan grid over a 2000 $\times$ 2000\,m area representing an urban region. Consecutive intersections are 500\,m apart, and each road has three lanes in each direction. Of the 25 intersections, the 21 that are not at the corners are controlled by fixed-cycle (40\,s) traffic lights. This regular grid topology provides repeatable line-of-sight conditions and regular intersection interactions, allowing the comparison to be conducted on a controlled basis.

Throughout the simulation, 150 vehicles move along predefined routes. As they traverse the grid, vehicles accelerate and decelerate according to traffic lights and vehicles ahead, producing a dynamic topology in which inter-vehicle neighborhood relationships change continuously over time. The total simulation time is 1000\,s.

A single RSU acts as the certificate authority (CA). It is placed at the center of the area, its coverage is not limited, and it is responsible for collecting trust feedback from vehicles, periodically computing global trust values, and applying identity revocation upon threshold violation. RSU coverage limits and multi-RSU coordination are not modeled in this study; since this choice applies identically to all compared methods, it isolates the observed performance differences from infrastructure effects, attributing them solely to the trust mechanisms.

Inter-vehicle communication is configured so that the set of event witnesses is a subset of the set of listeners: the event impact radius (250\,m) is kept smaller than the vehicles' broadcast range (500\,m). This distinguishes vehicles that directly observe an event from those that merely receive report messages, enabling a meaningful test of indirect trust evaluation. Report messages are forwarded over a single hop only. The environment and protocol parameters are summarized in Table~\ref{tab:sim_params}.

\begin{table}[!t]
\centering
\caption{Simulation and protocol parameters}
\label{tab:sim_params}
\footnotesize
\begin{tabular}{@{}>{\raggedright\arraybackslash}p{0.46\columnwidth}>{\raggedright\arraybackslash}p{0.48\columnwidth}@{}}
\toprule
\textbf{Parameter} & \textbf{Value} \\
\midrule
\multicolumn{2}{l}{\textit{Network and mobility}} \\
Simulation area & 2000 $\times$ 2000\,m \\
Grid structure & 5 $\times$ 5 Manhattan, 500\,m blocks \\
Lanes & 3 per direction (bidirectional) \\
Traffic lights & 21, 40\,s fixed cycle \\
Number of vehicles & 150 \\
Simulation time & 1000\,s \\
Number of RSUs & 1 (CA), unlimited coverage \\
\midrule
\multicolumn{2}{l}{\textit{Events and messaging}} \\
Number of events & 40 (Table~\ref{tab:events}) \\
Event impact radius & 250\,m \\
Broadcast range & 500\,m, single hop \\
Witness report interval & 5\,s \\
Distance threshold $D_{Th}$ & 800\,m \\
Time threshold $T_{Th}$ & half of the active duration \\
Base perception error $p_0$ & 0.1 \\
\midrule
\multicolumn{2}{l}{\textit{Trust management}} \\
Global trust update interval $\Delta$ & 50\,s \\
Consecutive rounds for revocation $N_r$ & 3 \\
Independent repetitions (seeds) & 10 \\
\bottomrule
\end{tabular}
\end{table}

\begin{figure}[!t]
\centering
\includegraphics[width=\columnwidth]{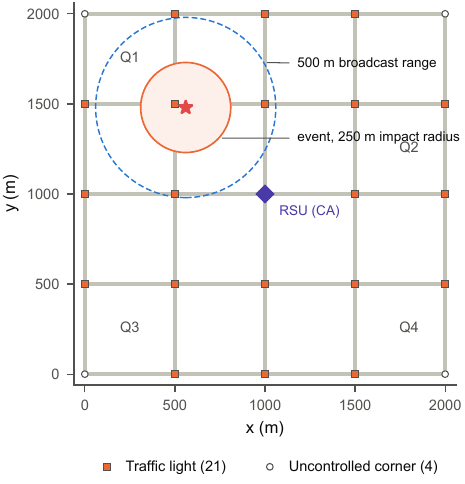}
\caption{Simulation scenario and network topology (road network from the SUMO configuration): a 5\,$\times$\,5 Manhattan grid with 21 traffic-light intersections, four quadrants, a central RSU acting as certificate authority, and an example event with its 250\,m impact radius (witnesses) and a 500\,m broadcast range around it (listeners).}
\label{fig:scenario}
\end{figure}

\subsection{Baselines and Variants}\label{subsec:baselines}
IPEK is compared against two baselines: TCEMD~\cite{tcemd}, a trust cascading-based emergency message dissemination model, and the beta reputation system~\cite{josang2002beta}. The primary reason for selecting TCEMD is architectural compatibility: like IPEK, it is centralized (an authority collects reports and computes global trust), updates global trust periodically, and applies threshold-based identity revocation. TCEMD is selected as a foundational trust-cascading approach for event-based trust systems.
In TCEMD, a witness rates each reporter with a binary feedback value, 1 when the report agrees with its perception and 0 otherwise. In each round, the CA computes the trust of a target as the reporter-trust-weighted fraction of positive feedback, scaled by activity weights derived from how often the target is rated and how often it rates others; the value is recomputed from the feedback of the round without accumulation. In the implementation used here, feedback is deduplicated per reporter, target and event within a round, and TCEMD's own revocation condition is replaced by the common revocation rule of Section~\ref{subsec:revocation}, so that the comparison concerns the trust computation only.

The beta reputation system~\cite{josang2002beta} is the standard probabilistic model for aggregating binary feedback and represents the natural alternative to an evidential fusion. It uses the same vehicle-side binary feedback as TCEMD. For each target $V_i$, the CA accumulates positive and negative evidence weighted by the reporter's trust at the start of the round, $r_i \leftarrow r_i + w F$ and $s_i \leftarrow s_i + w(1-F)$, and computes the trust as the expected value of the beta distribution, $(r_i + 1)/(r_i + s_i + 2)$, which starts at 0.5. Evidence is accumulated over the whole run without forgetting, feedback about oneself is ignored, and feedback is deduplicated within a round.

All methods are run with the same seeds. For a given seed, the vehicle mobility, the event schedule and the set of attackers are identical across methods; with revocation disabled (Section~\ref{subsec:protocol}), the report traffic and the number of feedback messages are identical as well, which was verified for every open-loop run. Differences between methods are therefore attributable to the trust computation alone.

Another recent DST-based method, DUEL~\cite{duel}, is excluded from the comparison for architectural reasons. DUEL is distributed: trust is maintained in pairwise relationships between neighbors, with no central authority, periodic global fusion, or identity revocation. Moreover, its reward/penalty signals are derived from message-drop/modification behavior rather than event-state reports. Consequently, DUEL's output semantics and evidence source cannot be meaningfully compared using this study's revocation-based metrics.

To quantify the contribution of IPEK's design components, five variants are defined, each differing from IPEK only in the relevant component. The \emph{Dempster} variant replaces the Yager combination rule used in IPEK's global trust fusion with the standard Dempster rule, leaving all other components unchanged. The standard Dempster rule redistributes conflicting mass by renormalizing with $(1-K)$, whereas the Yager rule transfers the same conflict to the uncertainty set; by changing only the conflict-handling strategy, this variant makes the effect of the combination rule visible. The \emph{NoSev} variant isolates the contribution of event-severity awareness. IPEK scales both the reward and penalty magnitudes in the local-trust update according to event severity; the NoSev variant replaces these two severity-dependent factors with constants chosen close to the scenario average (reward factor 0.55, penalty factor 0.80). Since the constants are chosen near the average, the variant differs from IPEK only in \emph{per-event} severity sensitivity, not in the average reward/penalty magnitude.
The remaining three variants remove one component each: \emph{NoCred} disables reporter credibility ($c_j \equiv 1$), \emph{NoRel} disables the observation-reliability factor ($\kappa_{ji} \equiv 1$), and \emph{NoRisk} disables risk accentuation ($\rho_{\text{risk}} = 0$). In the remainder of the paper, the Dempster variant is referred to as IPEK-Dempster.

\subsection{Ground Truth and Metrics}\label{subsec:metrics}
Performance is evaluated with four metrics widely used in the literature: detection rate (recall), precision, F1-score, and false positive rate (FPR), computed from the numbers of true positives ($TP$), false positives ($FP$), true negatives ($TN$), and false negatives ($FN$) by~\eqref{eq:recall}--\eqref{eq:fpr}:
\begin{align}
Recall &= \frac{TP}{TP + FN} \label{eq:recall} \\
Precision &= \frac{TP}{TP + FP} \label{eq:precision} \\
F1 &= \frac{2 \cdot Precision \cdot Recall}{Precision + Recall} \label{eq:f1eq} \\
FPR &= \frac{FP}{FP + TN} \label{eq:fpr}
\end{align}

The positive class consists of the vehicles designated as attackers at the start of a run. A revoked designated attacker is a true positive, including one revoked before it has launched an attack; a revoked vehicle that was never designated is a false positive, and the FPR is computed over the never-designated vehicles. This definition yields the same denominators for all methods within a run.
The detection delay of a revoked attacker is the time from its first attack opportunity, i.e., the first time its trigger condition holds while it witnesses an event, to its revocation; it is averaged over the attackers that were revoked after such an opportunity.

\subsection{Evaluation Protocol}\label{subsec:protocol}
In a closed loop, revocation feeds back into the trust computation: a revoked vehicle stops contributing reports and feedback, which changes the evidence available about every other vehicle. Each detection threshold therefore defines a separate experiment, and a receiver operating characteristic cannot be obtained from a single run. Moreover, the trust scores of the compared methods lie on different scales, so that a common numerical threshold range would favor some methods over others. The evaluation is therefore organized in two complementary layers.

\textbf{Open loop: separation without a threshold.} Revocation is disabled, and the global trust of every vehicle is recorded at the end of every round. For each vehicle, the score
\begin{equation}
s_i = \min_{k} \; \max\left\{ GT_i^{(k)}, \ldots, GT_i^{(k+N_r-1)} \right\}
\label{eq:score}
\end{equation}
is computed, where $GT_i^{(k)}$ is the global trust after round $k$ and the minimum runs over all windows of $N_r$ consecutive rounds. By construction, the revocation rule of Section~\ref{subsec:revocation} would be triggered for vehicle $V_i$ at threshold $\theta$ if and only if $s_i < \theta$; a vehicle with fewer than $N_r$ recorded rounds cannot be revoked and receives $s_i = +\infty$. Two measures are derived from the scores. The area under the receiver operating characteristic curve (AUC)~\cite{fawcett2006roc} is the probability that a randomly chosen attacker has a lower score than a randomly chosen honest vehicle, with ties counted as one half; it measures separation independently of any threshold and of the scale of the scores. The leave-one-seed-out F1 measures the separation achievable with a threshold chosen without access to the evaluated run: for each seed, the threshold is selected among all distinct scores observed in the other nine seeds of the same configuration so as to maximize their mean F1, with ties broken in favor of the lowest threshold, and the resulting threshold is applied to the held-out seed.

\textbf{Closed loop: deployment with a frozen threshold.} Revocation is enabled. For each method, the detection threshold is swept over eight values in a reference configuration, the dual-channel severity-triggered attack (D-Sev) with $\pi_a = 0.25$; the value with the highest mean F1 is selected and then kept fixed in all other configurations. The selection criterion was declared before the runs. The selected thresholds are 0.39 for IPEK (swept over 0.20--0.50), 0.70 for beta reputation (0.50--0.85), 0.60 for TCEMD (0.30--0.80) and $10^{-4}$ for IPEK-Dempster ($10^{-9}$--0.33); all lie in the interior of the swept ranges. Since the threshold is not re-tuned, the closed-loop results also measure how well the calibration of each method transfers across attacks and attacker ratios. The variants of Section~\ref{subsec:baselines} are swept in the same way in two reference configurations (D-Sev and R-Sev, $\pi_a = 0.25$) and compared with IPEK, each at its own best threshold. Operating points at matched FPR are obtained from these sweeps by linear interpolation between the operating points of each seed.

\textbf{Attack grid and statistics.} The dual-channel attacks are evaluated at nominal attacker ratios $\pi_a \in \{0.10, 0.25, 0.30, 0.35, 0.40\}$ and the report-only attacks at $\pi_a \in \{0.10, 0.25, 0.40\}$; every configuration is run with ten seeds, and the study comprises 4080 runs in total. Since the same seed yields the same attackers and the same mobility for every method, methods are compared with the paired two-sided Wilcoxon signed-rank test~\cite{wilcoxon1945} over the ten seeds; the smallest attainable p-value is 0.002. Results are reported as means, with 95\% confidence intervals based on the $t$ distribution where indicated. No correction for multiple comparisons is applied in general; where a claim rests on a family of comparisons, its robustness to the Holm correction~\cite{holm1979} is stated.

\textbf{Use of AI tools.} A large language model (Claude, Anthropic) was used to independently recompute the reported metrics and statistical tests from the raw simulation outputs, to check the consistency between the simulation code and the model description in Section~\ref{sec:ipek}, and for language editing. The design of the method, the simulation code, the experiments and all scientific decisions are the author's own; all AI-assisted content was reviewed and verified by the author.
\section{Results}\label{sec:results}
This section first compares IPEK with the baselines in terms of separation (open loop) and deployment performance (closed loop), then locates the attacker share at which the methods break down, and finally examines the contribution of each component, the combination rule, the sensitivity to the parameters and the communication overhead.

\subsection{Separation Without a Threshold}\label{subsec:open}
Table~\ref{tab:open} reports the open-loop AUC and the leave-one-seed-out F1 of the three methods for all attacks and attacker ratios.

\begin{sidewaystable*}
\centering
\caption{Open-loop separation: AUC (mean $\pm$ 95\% confidence interval) and leave-one-seed-out F1 (mean) over ten seeds. On a baseline value, $^{\dag}$ indicates that IPEK is significantly higher and $^{\ddag}$ that it is significantly lower (paired Wilcoxon, $p < 0.05$).}
\label{tab:open}
\footnotesize
\begin{tabular}{llcccccc}
\toprule
& & \multicolumn{3}{c}{\textbf{AUC}} & \multicolumn{3}{c}{\textbf{F1 (leave-one-seed-out)}} \\
\cmidrule(lr){3-5}\cmidrule(lr){6-8}
\textbf{Attack} & \textbf{$\pi_a$ (\%)} & \textbf{IPEK} & \textbf{Beta} & \textbf{TCEMD} & \textbf{IPEK} & \textbf{Beta} & \textbf{TCEMD} \\
\midrule
D-Sev & 10 & 0.976$\pm$0.022 & 0.916$\pm$0.049$^{\dag}$ & 0.814$\pm$0.062$^{\dag}$ & 0.933 & 0.617$^{\dag}$ & 0.442$^{\dag}$ \\
D-Sev & 25 & 0.904$\pm$0.053 & 0.785$\pm$0.060$^{\dag}$ & 0.670$\pm$0.069$^{\dag}$ & 0.752 & 0.546$^{\dag}$ & 0.455$^{\dag}$ \\
D-Sev & 30 & 0.801$\pm$0.097 & 0.682$\pm$0.077$^{\dag}$ & 0.573$\pm$0.074$^{\dag}$ & 0.655 & 0.467$^{\dag}$ & 0.463$^{\dag}$ \\
D-Sev & 35 & 0.560$\pm$0.187 & 0.524$\pm$0.097 & 0.437$\pm$0.083 & 0.538 & 0.538 & 0.537 \\
D-Sev & 40 & 0.366$\pm$0.188 & 0.415$\pm$0.099 & 0.341$\pm$0.068 & 0.584 & 0.586 & 0.585 \\
D-Loc & 10 & 0.993$\pm$0.012 & 0.961$\pm$0.029$^{\dag}$ & 0.874$\pm$0.054$^{\dag}$ & 0.946 & 0.741$^{\dag}$ & 0.538$^{\dag}$ \\
D-Loc & 25 & 0.959$\pm$0.020 & 0.852$\pm$0.055$^{\dag}$ & 0.730$\pm$0.050$^{\dag}$ & 0.871 & 0.678$^{\dag}$ & 0.521$^{\dag}$ \\
D-Loc & 30 & 0.880$\pm$0.070 & 0.727$\pm$0.079$^{\dag}$ & 0.605$\pm$0.066$^{\dag}$ & 0.769 & 0.572$^{\dag}$ & 0.481$^{\dag}$ \\
D-Loc & 35 & 0.614$\pm$0.204 & 0.531$\pm$0.103 & 0.454$\pm$0.083 & 0.498 & 0.538 & 0.538 \\
D-Loc & 40 & 0.375$\pm$0.229 & 0.384$\pm$0.090 & 0.325$\pm$0.072 & 0.585 & 0.586 & 0.586 \\
D-OnOff & 10 & 0.997$\pm$0.003 & 0.964$\pm$0.013$^{\dag}$ & 0.804$\pm$0.055$^{\dag}$ & 0.932 & 0.690$^{\dag}$ & 0.392$^{\dag}$ \\
D-OnOff & 25 & 0.876$\pm$0.056 & 0.856$\pm$0.043 & 0.677$\pm$0.059$^{\dag}$ & 0.699 & 0.652 & 0.480$^{\dag}$ \\
D-OnOff & 30 & 0.694$\pm$0.121 & 0.753$\pm$0.069 & 0.598$\pm$0.050 & 0.539 & 0.586 & 0.475 \\
D-OnOff & 35 & 0.415$\pm$0.187 & 0.585$\pm$0.097$^{\ddag}$ & 0.476$\pm$0.059 & 0.538 & 0.538 & 0.538 \\
D-OnOff & 40 & 0.205$\pm$0.149 & 0.446$\pm$0.107$^{\ddag}$ & 0.391$\pm$0.060$^{\ddag}$ & 0.586 & 0.586 & 0.586 \\
D-Cont & 10 & 1.000$\pm$0.000 & 0.999$\pm$0.002 & 0.958$\pm$0.018$^{\dag}$ & 0.997 & 0.976$^{\dag}$ & 0.749$^{\dag}$ \\
D-Cont & 25 & 0.995$\pm$0.005 & 0.985$\pm$0.013 & 0.855$\pm$0.061$^{\dag}$ & 0.963 & 0.922$^{\dag}$ & 0.683$^{\dag}$ \\
D-Cont & 30 & 0.935$\pm$0.054 & 0.918$\pm$0.064 & 0.650$\pm$0.141$^{\dag}$ & 0.854 & 0.810 & 0.529$^{\dag}$ \\
D-Cont & 35 & 0.535$\pm$0.289 & 0.646$\pm$0.184 & 0.354$\pm$0.201$^{\dag}$ & 0.401 & 0.579$^{\ddag}$ & 0.536$^{\ddag}$ \\
D-Cont & 40 & 0.295$\pm$0.270 & 0.385$\pm$0.208 & 0.170$\pm$0.122 & 0.586 & 0.586 & 0.583 \\
\midrule
R-Sev & 10 & 0.989$\pm$0.016 & 0.947$\pm$0.039$^{\dag}$ & 0.873$\pm$0.047$^{\dag}$ & 0.950 & 0.684$^{\dag}$ & 0.522$^{\dag}$ \\
R-Sev & 25 & 0.992$\pm$0.008 & 0.953$\pm$0.028$^{\dag}$ & 0.876$\pm$0.042$^{\dag}$ & 0.968 & 0.823$^{\dag}$ & 0.691$^{\dag}$ \\
R-Sev & 40 & 0.992$\pm$0.008 & 0.953$\pm$0.027$^{\dag}$ & 0.872$\pm$0.044$^{\dag}$ & 0.976 & 0.884$^{\dag}$ & 0.769$^{\dag}$ \\
R-Loc & 10 & 0.997$\pm$0.005 & 0.978$\pm$0.019$^{\dag}$ & 0.910$\pm$0.042$^{\dag}$ & 0.966 & 0.820$^{\dag}$ & 0.607$^{\dag}$ \\
R-Loc & 25 & 0.998$\pm$0.003 & 0.980$\pm$0.016$^{\dag}$ & 0.918$\pm$0.033$^{\dag}$ & 0.985 & 0.892$^{\dag}$ & 0.754$^{\dag}$ \\
R-Loc & 40 & 0.999$\pm$0.002 & 0.977$\pm$0.018$^{\dag}$ & 0.920$\pm$0.033$^{\dag}$ & 0.988 & 0.921$^{\dag}$ & 0.828$^{\dag}$ \\
R-OnOff & 10 & 1.000$\pm$0.001 & 0.986$\pm$0.005$^{\dag}$ & 0.861$\pm$0.043$^{\dag}$ & 0.980 & 0.795$^{\dag}$ & 0.471$^{\dag}$ \\
R-OnOff & 25 & 1.000$\pm$0.000 & 0.987$\pm$0.007$^{\dag}$ & 0.883$\pm$0.033$^{\dag}$ & 0.993 & 0.896$^{\dag}$ & 0.698$^{\dag}$ \\
R-OnOff & 40 & 1.000$\pm$0.001 & 0.987$\pm$0.008$^{\dag}$ & 0.881$\pm$0.026$^{\dag}$ & 0.993 & 0.941$^{\dag}$ & 0.790$^{\dag}$ \\
R-Cont & 10 & 1.000$\pm$0.000 & 0.999$\pm$0.001 & 0.972$\pm$0.012$^{\dag}$ & 0.996 & 0.984 & 0.797$^{\dag}$ \\
R-Cont & 25 & 1.000$\pm$0.000 & 0.999$\pm$0.001 & 0.975$\pm$0.013$^{\dag}$ & 0.999 & 0.992 & 0.902$^{\dag}$ \\
R-Cont & 40 & 1.000$\pm$0.000 & 0.999$\pm$0.002 & 0.977$\pm$0.013$^{\dag}$ & 0.999 & 0.993$^{\dag}$ & 0.949$^{\dag}$ \\
\bottomrule
\end{tabular}
\end{sidewaystable*}

\textbf{Up to 25\% attackers.} In the 16 configurations with $\pi_a \leq 0.25$, IPEK separates attackers from honest vehicles significantly better than TCEMD in all 16 and better than beta reputation in 11, and it is not significantly worse than either baseline in any of them. The five configurations without a significant difference from beta reputation are the continuous attacks (D-Cont and R-Cont at 10\% and 25\%), in which both methods are close to perfect separation (AUC $\geq 0.985$), and the dual-channel on-off attack at 25\% (0.876 versus 0.856). The leave-one-seed-out F1 confirms this picture: IPEK is significantly higher than TCEMD in all 16 configurations and higher than beta reputation in 13, with no configuration in which it is lower. The largest margins occur under the attacks that IPEK is designed for: under D-Sev and D-Loc at 25\%, the AUC of IPEK is 0.904 and 0.959, compared with 0.785 and 0.852 for beta reputation and 0.670 and 0.730 for TCEMD.

\textbf{Report-only attacks.} When attackers falsify only their reports, IPEK attains an AUC of at least 0.989 at every attacker ratio, including 40\%, and its leave-one-seed-out F1 is at least 0.950. Beta reputation remains below IPEK in all severity-, location- and on-off-triggered configurations (AUC 0.947--0.987), and TCEMD remains below both. Under the continuous report-only attack, both IPEK and beta reputation separate almost perfectly. The report channel alone, without the support of corrupted feedback, is therefore not sufficient for attackers to hide among honest vehicles, even when they make up 40\% of the population.

\textbf{Higher attacker ratios under dual-channel attacks.} At 30\%, IPEK retains a significant advantage over beta reputation under the severity- and location-triggered attacks (AUC 0.801 versus 0.682 and 0.880 versus 0.727), whereas the differences under the on-off and continuous attacks are not significant. At 35\% and 40\%, separation deteriorates for all three methods: the AUC values lie between 0.17 and 0.65, and many runs fall below 0.5, i.e., attackers receive higher scores than honest vehicles. In this region, beta reputation degrades more gracefully than IPEK under the on-off attack, where IPEK is significantly lower at 35\% and 40\%. The leave-one-seed-out F1 values of about 0.58 at 40\% correspond to revoking nearly every vehicle and do not indicate a usable operating point for any method. Section~\ref{subsec:break} locates the breakdown more precisely.

Over all 32 configurations, the AUC of IPEK is significantly higher than that of beta reputation in 16, lower in 2 (D-OnOff at 35\% and 40\%) and not different in 14; compared with TCEMD, it is higher in 24, lower in 1 and not different in 7.

\subsection{Deployment with a Frozen Threshold}\label{subsec:closed}
\begin{figure*}[!t]
\centering
\includegraphics[width=\textwidth]{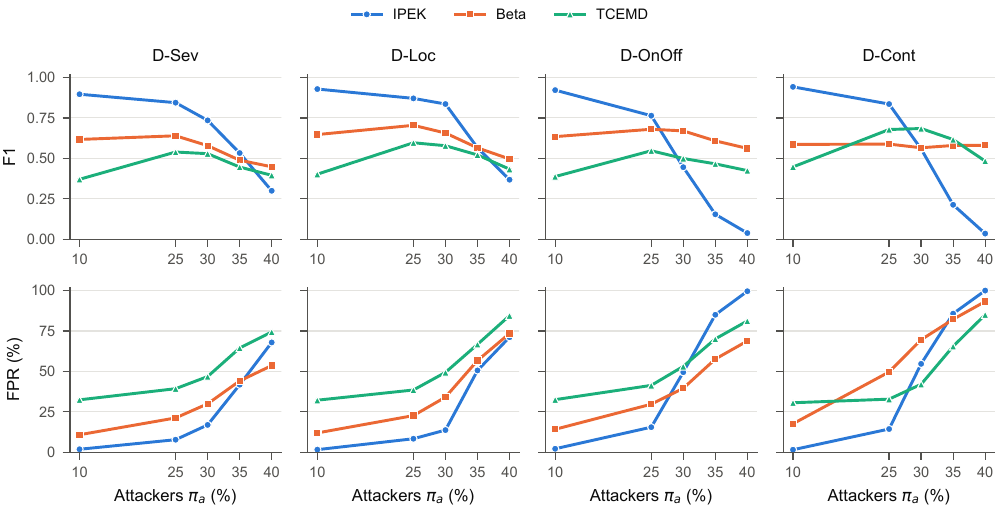}
\caption{Closed-loop F1 (top) and FPR (bottom) with frozen thresholds under the four dual-channel attacks, as functions of the nominal attacker ratio (means over ten seeds; confidence intervals in Table~\ref{tab:closed}).}
\label{fig:closed}
\end{figure*}

\begin{figure}[!t]
\centering
\includegraphics[width=\columnwidth]{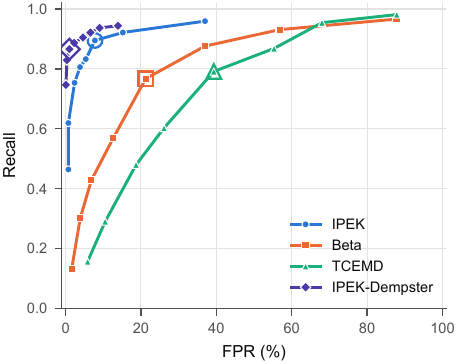}
\caption{Operating points of the four methods in the reference configuration (D-Sev, $\pi_a = 0.25$) over their threshold sweeps (mean recall and FPR over ten seeds per threshold). The large hollow marker indicates the frozen threshold of each method.}
\label{fig:operating}
\end{figure}

Table~\ref{tab:closed} reports the closed-loop results with the thresholds selected in the reference configuration and frozen thereafter, and Fig.~\ref{fig:closed} shows the F1 and FPR under the dual-channel attacks as functions of the attacker ratio. Fig.~\ref{fig:operating} shows the operating points that each method can reach in the reference configuration when its threshold is varied: IPEK reaches a recall of 0.8 at an FPR of about 4\%, whereas beta reputation and TCEMD need about 26\% and 41\%, respectively.

\begin{sidewaystable*}
\centering
\caption{Closed-loop performance with frozen thresholds (IPEK 0.39, beta reputation 0.70, TCEMD 0.60): F1 (mean $\pm$ 95\% confidence interval), FPR, recall and detection delay, means over ten seeds. For F1, $^{\dag}$ ($^{\ddag}$) on a baseline value indicates that IPEK is significantly higher (lower); for the delay, $^{\dag}$ ($^{\ddag}$) indicates that IPEK is significantly faster (slower) (paired Wilcoxon, $p < 0.05$).}
\label{tab:closed}
\scriptsize
\setlength{\tabcolsep}{3pt}
\begin{tabular}{llcccccccccccc}
\toprule
& & \multicolumn{3}{c}{\textbf{F1}} & \multicolumn{3}{c}{\textbf{FPR (\%)}} & \multicolumn{3}{c}{\textbf{Recall}} & \multicolumn{3}{c}{\textbf{Delay (s)}} \\
\cmidrule(lr){3-5}\cmidrule(lr){6-8}\cmidrule(lr){9-11}\cmidrule(lr){12-14}
\textbf{Attack} & \textbf{$\pi_a$ (\%)} & \textbf{IPEK} & \textbf{Beta} & \textbf{TCEMD} & \textbf{IPEK} & \textbf{Beta} & \textbf{TCEMD} & \textbf{IPEK} & \textbf{Beta} & \textbf{TCEMD} & \textbf{IPEK} & \textbf{Beta} & \textbf{TCEMD} \\
\midrule
D-Sev & 10 & 0.897$\pm$0.030 & 0.617$\pm$0.076$^{\dag}$ & 0.371$\pm$0.059$^{\dag}$ & 1.9 & 10.8 & 32.4 & 0.938 & 0.850 & 0.825 & 187 & 193 & 246$^{\dag}$ \\
D-Sev & 25 & 0.845$\pm$0.040 & 0.640$\pm$0.039$^{\dag}$ & 0.539$\pm$0.065$^{\dag}$ & 7.8 & 21.3 & 39.3 & 0.895 & 0.767 & 0.791 & 251 & 207$^{\ddag}$ & 256 \\
D-Sev & 30 & 0.735$\pm$0.120 & 0.579$\pm$0.046$^{\dag}$ & 0.529$\pm$0.074$^{\dag}$ & 16.9 & 30.0 & 46.8 & 0.786 & 0.685 & 0.738 & 290 & 200$^{\ddag}$ & 267 \\
D-Sev & 35 & 0.533$\pm$0.198 & 0.489$\pm$0.080 & 0.447$\pm$0.066 & 41.8 & 44.1 & 64.5 & 0.593 & 0.564 & 0.600 & 273 & 187$^{\ddag}$ & 271 \\
D-Sev & 40 & 0.300$\pm$0.145 & 0.448$\pm$0.080$^{\ddag}$ & 0.395$\pm$0.058 & 67.9 & 53.6 & 74.4 & 0.351 & 0.512 & 0.505 & 247 & 182$^{\ddag}$ & 294 \\
D-Loc & 10 & 0.929$\pm$0.042 & 0.647$\pm$0.046$^{\dag}$ & 0.402$\pm$0.065$^{\dag}$ & 1.6 & 12.0 & 32.2 & 0.988 & 0.952 & 0.908 & 196 & 154$^{\ddag}$ & 201 \\
D-Loc & 25 & 0.871$\pm$0.049 & 0.705$\pm$0.046$^{\dag}$ & 0.597$\pm$0.061$^{\dag}$ & 8.4 & 22.7 & 38.5 & 0.953 & 0.906 & 0.899 & 239 & 193$^{\ddag}$ & 242 \\
D-Loc & 30 & 0.836$\pm$0.066 & 0.657$\pm$0.064$^{\dag}$ & 0.578$\pm$0.074$^{\dag}$ & 13.7 & 34.1 & 49.4 & 0.928 & 0.860 & 0.853 & 272 & 202$^{\ddag}$ & 270 \\
D-Loc & 35 & 0.566$\pm$0.226 & 0.565$\pm$0.072 & 0.521$\pm$0.079 & 50.6 & 56.7 & 66.7 & 0.668 & 0.768 & 0.753 & 272 & 197$^{\ddag}$ & 276 \\
D-Loc & 40 & 0.368$\pm$0.231 & 0.497$\pm$0.080 & 0.433$\pm$0.062 & 71.2 & 73.5 & 84.5 & 0.449 & 0.678 & 0.613 & 249 & 186$^{\ddag}$ & 274 \\
D-OnOff & 10 & 0.922$\pm$0.034 & 0.635$\pm$0.073$^{\dag}$ & 0.388$\pm$0.050$^{\dag}$ & 2.2 & 14.2 & 32.5 & 1.000 & 0.989 & 0.877 & 334 & 183$^{\ddag}$ & 326 \\
D-OnOff & 25 & 0.764$\pm$0.050 & 0.680$\pm$0.033$^{\dag}$ & 0.547$\pm$0.042$^{\dag}$ & 15.5 & 29.7 & 41.4 & 0.888 & 0.956 & 0.831 & 445 & 199$^{\ddag}$ & 342$^{\ddag}$ \\
D-OnOff & 30 & 0.446$\pm$0.191 & 0.670$\pm$0.034$^{\ddag}$ & 0.500$\pm$0.046 & 49.5 & 39.6 & 53.1 & 0.552 & 0.941 & 0.733 & 422 & 210$^{\ddag}$ & 336$^{\ddag}$ \\
D-OnOff & 35 & 0.154$\pm$0.156 & 0.609$\pm$0.067$^{\ddag}$ & 0.466$\pm$0.048$^{\ddag}$ & 85.0 & 57.7 & 70.1 & 0.191 & 0.857 & 0.668 & 371 & 203$^{\ddag}$ & 344 \\
D-OnOff & 40 & 0.038$\pm$0.035 & 0.562$\pm$0.081$^{\ddag}$ & 0.425$\pm$0.055$^{\ddag}$ & 99.6 & 68.8 & 81.2 & 0.053 & 0.761 & 0.578 & 285 & 199 & 340 \\
D-Cont & 10 & 0.942$\pm$0.033 & 0.587$\pm$0.049$^{\dag}$ & 0.447$\pm$0.062$^{\dag}$ & 1.6 & 17.6 & 30.6 & 1.000 & 1.000 & 1.000 & 226 & 127$^{\ddag}$ & 140$^{\ddag}$ \\
D-Cont & 25 & 0.836$\pm$0.036 & 0.589$\pm$0.036$^{\dag}$ & 0.679$\pm$0.049$^{\dag}$ & 14.4 & 49.7 & 32.9 & 1.000 & 1.000 & 1.000 & 268 & 130$^{\ddag}$ & 152$^{\ddag}$ \\
D-Cont & 30 & 0.558$\pm$0.204 & 0.566$\pm$0.020 & 0.685$\pm$0.043 & 54.7 & 69.6 & 42.0 & 0.779 & 0.998 & 1.000 & 304 & 131$^{\ddag}$ & 165$^{\ddag}$ \\
D-Cont & 35 & 0.213$\pm$0.216 & 0.580$\pm$0.026$^{\ddag}$ & 0.617$\pm$0.088$^{\ddag}$ & 85.7 & 82.2 & 65.5 & 0.288 & 0.976 & 0.917 & 233 & 131$^{\ddag}$ & 190 \\
D-Cont & 40 & 0.035$\pm$0.037 & 0.581$\pm$0.022$^{\ddag}$ & 0.484$\pm$0.124$^{\ddag}$ & 100.0 & 93.2 & 85.1 & 0.050 & 0.948 & 0.707 & 171 & 135 & 178 \\
\midrule
R-Sev & 10 & 0.918$\pm$0.033 & 0.689$\pm$0.079$^{\dag}$ & 0.414$\pm$0.062$^{\dag}$ & 1.3 & 7.8 & 29.1 & 0.946 & 0.868 & 0.877 & 179 & 169 & 222$^{\dag}$ \\
R-Sev & 25 & 0.960$\pm$0.019 & 0.826$\pm$0.055$^{\dag}$ & 0.653$\pm$0.061$^{\dag}$ & 0.9 & 7.9 & 27.0 & 0.948 & 0.873 & 0.867 & 184 & 176 & 207$^{\dag}$ \\
R-Sev & 40 & 0.972$\pm$0.014 & 0.869$\pm$0.053$^{\dag}$ & 0.749$\pm$0.057$^{\dag}$ & 0.7 & 7.9 & 28.0 & 0.955 & 0.859 & 0.840 & 185 & 182 & 211$^{\dag}$ \\
R-Loc & 10 & 0.952$\pm$0.031 & 0.730$\pm$0.066$^{\dag}$ & 0.428$\pm$0.065$^{\dag}$ & 1.0 & 8.1 & 30.3 & 0.987 & 0.957 & 0.940 & 184 & 148$^{\ddag}$ & 187 \\
R-Loc & 25 & 0.979$\pm$0.010 & 0.871$\pm$0.043$^{\dag}$ & 0.688$\pm$0.067$^{\dag}$ & 0.9 & 8.0 & 27.5 & 0.984 & 0.957 & 0.943 & 178 & 161$^{\ddag}$ & 191 \\
R-Loc & 40 & 0.988$\pm$0.007 & 0.921$\pm$0.029$^{\dag}$ & 0.812$\pm$0.041$^{\dag}$ & 0.8 & 8.2 & 27.6 & 0.987 & 0.953 & 0.952 & 179 & 163$^{\ddag}$ & 206$^{\dag}$ \\
R-OnOff & 10 & 0.962$\pm$0.027 & 0.731$\pm$0.069$^{\dag}$ & 0.408$\pm$0.057$^{\dag}$ & 1.0 & 8.7 & 30.1 & 1.000 & 0.989 & 0.880 & 280 & 173$^{\ddag}$ & 312 \\
R-OnOff & 25 & 0.982$\pm$0.008 & 0.878$\pm$0.041$^{\dag}$ & 0.672$\pm$0.052$^{\dag}$ & 1.0 & 8.7 & 26.0 & 0.993 & 0.986 & 0.892 & 277 & 172$^{\ddag}$ & 297 \\
R-OnOff & 40 & 0.990$\pm$0.009 & 0.938$\pm$0.021$^{\dag}$ & 0.783$\pm$0.030$^{\dag}$ & 0.8 & 8.3 & 28.2 & 0.991 & 0.986 & 0.898 & 283 & 168$^{\ddag}$ & 302 \\
R-Cont & 10 & 0.949$\pm$0.030 & 0.737$\pm$0.064$^{\dag}$ & 0.464$\pm$0.068$^{\dag}$ & 1.3 & 8.7 & 28.5 & 1.000 & 1.000 & 1.000 & 208 & 126$^{\ddag}$ & 138$^{\ddag}$ \\
R-Cont & 25 & 0.983$\pm$0.011 & 0.880$\pm$0.039$^{\dag}$ & 0.718$\pm$0.052$^{\dag}$ & 1.2 & 9.2 & 26.9 & 1.000 & 1.000 & 1.000 & 205 & 127$^{\ddag}$ & 140$^{\ddag}$ \\
R-Cont & 40 & 0.996$\pm$0.003 & 0.943$\pm$0.019$^{\dag}$ & 0.850$\pm$0.035$^{\dag}$ & 0.6 & 8.6 & 25.0 & 1.000 & 1.000 & 1.000 & 205 & 126$^{\ddag}$ & 142$^{\ddag}$ \\
\bottomrule
\end{tabular}
\end{sidewaystable*}

\textbf{Up to 25\% attackers.} With $\pi_a \leq 0.25$, IPEK achieves a significantly higher F1 than both baselines in all 16 configurations ($p \leq 0.027$); within each family of 16 comparisons, all differences remain significant after the Holm correction. The F1 of IPEK ranges from 0.764 to 0.983, against 0.587--0.880 for beta reputation and 0.371--0.718 for TCEMD, and the confidence half-widths of IPEK do not exceed 0.05. The difference is driven mainly by false positives: the FPR of IPEK lies between 0.9\% and 15.5\%, whereas that of beta reputation lies between 7.8\% and 49.7\% and that of TCEMD between 26.0\% and 41.4\%. None of the 160 runs of IPEK in this region exceeds an FPR of 30\%. Under the report-only attacks, IPEK keeps its mean FPR below 1.4\% at every attacker ratio and reaches an F1 between 0.972 and 0.996 at 40\%.

The closed-loop margins should be read together with the open-loop results. Under the continuous attacks (D-Cont and R-Cont), the open-loop separation of IPEK and beta reputation is statistically indistinguishable, yet the closed-loop F1 of beta reputation is markedly lower, e.g., 0.589 versus 0.836 under D-Cont at 25\%. In these configurations, the frozen threshold of 0.70 cuts into the trust distribution of honest vehicles under beta reputation, producing an FPR between 8.7\% and 49.7\%; the effect is strongest under D-Cont, where bad-mouthing pushes the trust of honest vehicles downward. The closed-loop advantage of IPEK in these four configurations therefore reflects a better transfer of its calibration across attacks rather than a better separation. In the severity- and location-triggered attacks, by contrast, IPEK is superior in both layers.

\textbf{Higher attacker ratios.} At 30\%, IPEK remains significantly better than both baselines under D-Sev (F1 0.735 versus 0.579 and 0.529) and D-Loc (0.836 versus 0.657 and 0.578). Under D-Cont, the three methods are not significantly different, and under D-OnOff IPEK is below beta reputation (0.446 versus 0.670, $p = 0.049$). At 35\% and 40\%, the failure mode of IPEK differs from that of the baselines. IPEK inverts: honest vehicles are revoked while attackers survive, with recall between 0.05 and 0.67 and FPR between 42\% and 100\%. Beta reputation and TCEMD revoke indiscriminately, with FPR between 44\% and 93\% and between 65\% and 85\%, respectively; their F1 values of 0.40--0.62 are a consequence of high recall at high FPR, not of separation. None of the three methods offers a usable operating point in this region.

\textbf{Detection delay.} IPEK revokes attackers later than beta reputation: with $\pi_a \leq 0.25$, beta reputation is significantly faster in 13 of the 16 configurations, and IPEK is faster in none. The mean delays range from 178\,s to 445\,s for IPEK, from 126\,s to 207\,s for beta reputation and from 138\,s to 342\,s for TCEMD. The faster reaction of beta reputation comes with its higher FPR. The delay of IPEK is a design setting rather than a fixed property: with $N_r = 1$ instead of 3, the delay under D-Sev at 25\% drops from 251\,s to 137\,s at the cost of an FPR of 21.4\% (Section~\ref{subsec:sens}).

\textbf{Effect of revocation.} Under dual-channel attacks with $\pi_a \leq 0.25$, the closed-loop F1 of IPEK is higher than the open-loop F1 obtained with the same threshold in all eight configurations, significantly in five: revoking an attacker removes its corrupted feedback and protects the evidence about the remaining vehicles. The same holds at 30\% under D-Sev and D-Loc. Under the continuous attack at 30\%, revocation has the opposite effect (F1 0.558 versus 0.702, $p = 0.008$): erroneous revocations of honest vehicles reduce the number of honest reporters and accelerate the degradation; under the on-off attack at 30\%, the effect points in the same direction but is not significant.

Over all 32 configurations, IPEK achieves a significantly higher F1 than beta reputation in 22, a lower one in 6 and an equal one in 4; compared with TCEMD, the counts are 22, 4 and 6.

\subsection{Breakdown Point}\label{subsec:break}
\begin{figure}[!t]
\centering
\includegraphics[width=\columnwidth]{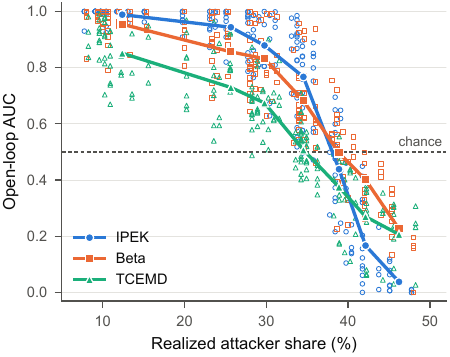}
\caption{Open-loop AUC of every run of the four dual-channel attacks against the realized share of attackers (hollow markers), with the mean per share interval of Table~\ref{tab:break} (lines). Values below 0.5 indicate that attackers receive higher scores than honest vehicles.}
\label{fig:break}
\end{figure}

Since attackers are assigned at random, the realized share of attackers differs from the nominal ratio, and runs with the same nominal ratio may lie on either side of the breakdown point. Table~\ref{tab:break} therefore groups all runs of the four dual-channel attacks by their realized attacker share, and Fig.~\ref{fig:break} shows the open-loop AUC of every run against it.

\begin{table*}[!t]
\centering
\caption{Performance under dual-channel attacks grouped by the realized share of attackers (all runs of the four dual-channel attacks, 200 per method).}
\label{tab:break}
\footnotesize
\setlength{\tabcolsep}{3pt}
\begin{tabular}{lccccccc}
\toprule
Realized attacker share (\%) & $\leq$20 & 21--28 & 29--32 & 33--36 & 37--40 & 41--44 & $>$44 \\
\midrule
Runs per method & 48 & 32 & 24 & 40 & 24 & 20 & 12 \\
\midrule
\multicolumn{8}{l}{\textit{Open loop: mean AUC [runs with AUC $<$ 0.5]}} \\
IPEK & 0.99 [0] & 0.94 [0] & 0.88 [0] & 0.77 [2] & 0.44 [15] & 0.17 [20] & 0.04 [12] \\
Beta & 0.95 [0] & 0.86 [0] & 0.83 [0] & 0.68 [1] & 0.50 [11] & 0.40 [18] & 0.23 [12] \\
TCEMD & 0.85 [0] & 0.73 [1] & 0.67 [0] & 0.51 [18] & 0.38 [23] & 0.27 [20] & 0.21 [12] \\
IPEK-Dempster & 0.99 [0] & 0.91 [2] & 0.87 [3] & 0.52 [17] & 0.26 [18] & 0.05 [20] & 0.01 [12] \\
\midrule
\multicolumn{8}{l}{\textit{Closed loop, frozen threshold: mean F1 / mean FPR (\%)}} \\
IPEK & 0.91 / 3 & 0.83 / 12 & 0.76 / 19 & 0.49 / 53 & 0.29 / 73 & 0.12 / 94 & 0.04 / 99 \\
Beta & 0.63 / 15 & 0.63 / 33 & 0.65 / 36 & 0.61 / 51 & 0.55 / 66 & 0.52 / 74 & 0.44 / 86 \\
TCEMD & 0.42 / 33 & 0.58 / 40 & 0.61 / 41 & 0.55 / 58 & 0.52 / 68 & 0.43 / 85 & 0.34 / 95 \\
IPEK-Dempster & 0.98 / 0 & 0.86 / 8 & 0.80 / 13 & 0.42 / 50 & 0.21 / 69 & 0.04 / 92 & 0.01 / 97 \\
\bottomrule
\end{tabular}
\end{table*}

In the open loop, IPEK does not produce a single inverted run (AUC $< 0.5$) up to a realized share of 32\% (104 runs), and its mean AUC remains above that of both baselines up to 36\%. Beyond about one third, separation collapses for all methods, and IPEK collapses more sharply than beta reputation, which degrades more gradually. The closed loop shows the same boundary: up to 32\%, IPEK keeps a mean F1 of 0.76 or more at a mean FPR of at most 19\%, while between 33\% and 36\% its mean FPR already exceeds 50\%. The operating envelope of IPEK under fully colluding dual-channel attacks is therefore a realized attacker share of about one third. Under report-only attacks, no breakdown occurs up to the largest ratio considered (Section~\ref{subsec:open}).

The mechanism of the breakdown is the same for all evidence-aggregating methods: once colluding attackers form the majority among the witnesses of a target, their feedback defines the consensus. In IPEK, the reporter-credibility test then penalizes the honest minority instead of the attackers (Proposition~\ref{prop:cred}), and the global trust of honest vehicles is driven down, which explains the inversion observed in both layers.

\subsection{Component Analysis}\label{subsec:ablation}
Table~\ref{tab:abl_open} reports the open-loop effect of removing each component at 25\%, and Table~\ref{tab:abl_closed} compares each variant with IPEK at its own best threshold in the two reference configurations.

\begin{table*}[!t]
\centering
\caption{Open-loop component analysis at $\pi_a = 0.25$: AUC / leave-one-seed-out F1 (means over ten seeds). $^{*}$: significantly different from IPEK (paired Wilcoxon, $p < 0.05$).}
\label{tab:abl_open}
\scriptsize
\setlength{\tabcolsep}{2.5pt}
\begin{tabular}{lcccccc}
\toprule
\textbf{Attack} & \textbf{IPEK} & \textbf{NoSev} & \textbf{NoCred} & \textbf{NoRel} & \textbf{NoRisk} & \textbf{IPEK-Dempster} \\
\midrule
D-Sev & 0.904 / 0.752 & 0.843$^{*}$ / 0.686$^{*}$ & 0.529$^{*}$ / 0.407$^{*}$ & 0.907 / 0.771 & 0.896$^{*}$ / 0.729$^{*}$ & 0.961$^{*}$ / 0.903$^{*}$ \\
D-Loc & 0.959 / 0.871 & 0.930$^{*}$ / 0.836$^{*}$ & 0.551$^{*}$ / 0.392$^{*}$ & 0.962 / 0.887 & 0.952$^{*}$ / 0.865 & 0.988$^{*}$ / 0.947$^{*}$ \\
D-OnOff & 0.876 / 0.699 & 0.913$^{*}$ / 0.741 & 0.374$^{*}$ / 0.407$^{*}$ & 0.908$^{*}$ / 0.733$^{*}$ & 0.871$^{*}$ / 0.689 & 0.781 / 0.722 \\
D-Cont & 0.995 / 0.963 & 0.997$^{*}$ / 0.970 & 0.514$^{*}$ / 0.407$^{*}$ & 0.996 / 0.974$^{*}$ & 0.995 / 0.967 & 0.999 / 0.983$^{*}$ \\
\midrule
R-Sev & 0.992 / 0.968 & 0.981$^{*}$ / 0.929 & 0.991 / 0.964 & 0.989$^{*}$ / 0.953$^{*}$ & 0.991 / 0.962 & 0.989 / 0.958 \\
R-Loc & 0.998 / 0.985 & 0.997 / 0.966 & 0.998 / 0.980 & 0.998 / 0.976$^{*}$ & 0.998 / 0.981 & 0.996 / 0.985 \\
R-OnOff & 1.000 / 0.993 & 1.000 / 0.988 & 1.000 / 0.991 & 0.999$^{*}$ / 0.973$^{*}$ & 1.000 / 0.993 & 1.000 / 0.992 \\
R-Cont & 1.000 / 0.999 & 1.000 / 0.998 & 1.000 / 0.999 & 1.000 / 0.999 & 1.000 / 0.999 & 1.000 / 0.999 \\
\bottomrule
\end{tabular}
\end{table*}

\begin{table*}[!t]
\centering
\caption{Closed-loop component analysis in the reference configurations ($\pi_a = 0.25$): each variant at its own best threshold, means over ten seeds. $\Delta$F1 = F1(IPEK) $-$ F1(variant); $p$: paired Wilcoxon. NoCred is not swept under R-Sev, since its feedback is honest.}
\label{tab:abl_closed}
\footnotesize
\setlength{\tabcolsep}{3pt}
\begin{tabular}{llccccccc}
\toprule
\textbf{Attack} & \textbf{Variant} & \textbf{Threshold} & \textbf{F1} & \textbf{Recall} & \textbf{FPR (\%)} & \textbf{Delay (s)} & \textbf{$\Delta$F1} & \textbf{$p$} \\
\midrule
D-Sev & IPEK & 0.39 & 0.845 & 0.895 & 7.8 & 251 & -- & -- \\
 & NoSev & 0.4 & 0.745 & 0.707 & 5.6 & 315 & +0.099 & 0.020 \\
 & NoCred & 0.36 & 0.432 & 0.603 & 43.1 & 304 & +0.412 & 0.002 \\
 & NoRel & 0.28 & 0.843 & 0.853 & 5.8 & 289 & +0.002 & 0.625 \\
 & NoRisk & 0.36 & 0.833 & 0.817 & 4.7 & 292 & +0.011 & 0.695 \\
 & IPEK-Dempster & $10^{-4}$ & 0.911 & 0.866 & 1.0 & 289 & $-$0.066 & 0.020 \\
\midrule
R-Sev & IPEK & 0.39 & 0.960 & 0.948 & 0.9 & 184 & -- & -- \\
 & NoSev & 0.43 & 0.937 & 0.915 & 1.2 & 197 & +0.023 & 0.109 \\
 & NoRel & 0.28 & 0.952 & 0.938 & 1.1 & 193 & +0.009 & 0.219 \\
 & NoRisk & 0.39 & 0.956 & 0.939 & 0.8 & 186 & +0.005 & 0.219 \\
 & IPEK-Dempster & $10^{-3}$ & 0.955 & 0.920 & 0.1 & 199 & +0.005 & 0.770 \\
\bottomrule
\end{tabular}
\end{table*}

\textbf{Reporter credibility.} Without reporter credibility, IPEK cannot withstand dual-channel attacks: its AUC drops to 0.37--0.55 under the four dual-channel attacks, i.e., to chance level or below, and its closed-loop F1 in the reference configuration falls from 0.845 to 0.432 at an FPR of 43.1\%. Under report-only attacks, where feedback is honest, credibility has no measurable cost (AUC 0.991--1.000 with and without it). The component is thus indispensable against feedback manipulation and neutral otherwise. Section~\ref{subsec:sens} shows that it also makes IPEK insensitive to the strength of the manipulation.

\textbf{Criticality awareness.} The benefit of scaling rewards and penalties by event and location criticality depends on whether the attack follows criticality. Under the severity- and location-triggered dual-channel attacks, removing it lowers the AUC by 0.061 and 0.029, and in the closed loop it lowers the F1 by 0.099 at the respective best thresholds ($p = 0.020$). At matched FPR, the recall gain under D-Sev is 0.16 at 2\% FPR (0.745 versus 0.589, $p = 0.004$), 0.15 at 5\% and 0.11 at 10\% ($p = 0.008$), and attackers are revoked 26--43\,s earlier ($p \leq 0.037$). Under R-Sev, the recall gain at matched FPR is 0.054 at 0.5\% FPR and 0.040 at 1\% ($p \leq 0.027$). Under the on-off attack, which ignores criticality, the component has a cost: the AUC is 0.037 lower with it than without it. Under the continuous attacks, the effect is negligible. Criticality awareness is therefore a trade-off that pays off precisely against the strategic attackers it targets, and its cost when the attack is unrelated to criticality is small but measurable.

\textbf{Risk accentuation.} Removing risk accentuation lowers the AUC slightly but significantly under D-Sev, D-Loc and D-OnOff (by 0.005--0.008); at the respective best thresholds, the closed-loop F1 difference is not significant. At matched FPR under D-Sev, risk accentuation leads to earlier revocation, by 11\,s at 2\% FPR, 21\,s at 5\% and 8\,s at 10\% ($p = 0.014$--$0.039$), i.e., by 3--8\% of the delay, without a loss in recall. These gains are small and do not remain significant after a Holm correction over the six matched comparisons; risk accentuation is therefore best regarded as a minor refinement that slightly accelerates revocation.

\textbf{Observation reliability.} The observation-reliability factor yields small but significant open-loop gains under the report-only attacks (F1 higher by 0.009--0.020 under R-Sev, R-Loc and R-OnOff), no gain under the dual-channel attacks, and losses under D-OnOff (AUC 0.032 lower with the factor) and D-Cont (F1 0.011 lower), and no significant F1 difference at the best thresholds in the closed loop. At matched FPR under R-Sev, it is associated with earlier revocation at 0.5\% FPR (14\,s, $p = 0.004$); at this FPR level, however, the operating points correspond to fractions of a single honest vehicle per run, and the comparison is sensitive to the interpolation. The factor is retained in the model, since it addresses perception errors by construction, but no consistent performance benefit is claimed for it.

\subsection{Combination Rule}\label{subsec:rule}
\begin{figure*}[!t]
\centering
\includegraphics[width=\textwidth]{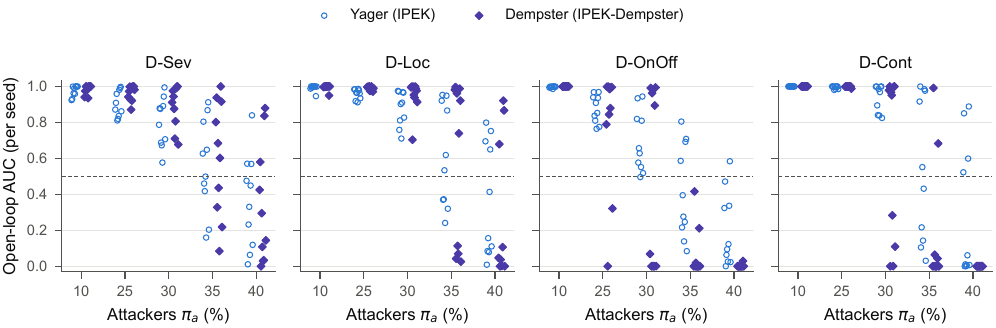}
\caption{Open-loop AUC of every seed under the dual-channel attacks with Yager's rule (IPEK, hollow circles) and Dempster's rule (IPEK-Dempster, filled diamonds). Dempster's rule concentrates its runs near 1 or near 0, whereas Yager's rule degrades more gradually.}
\label{fig:rule}
\end{figure*}

Table~\ref{tab:rule} compares Yager's rule, used in IPEK, with Dempster's rule under the dual-channel attacks, in which conflicting evidence is most pronounced, and Fig.~\ref{fig:rule} shows the open-loop AUC of every seed.

 Under the report-only attacks, the two rules are equivalent in the open loop (AUC differences of at most 0.005, none significant).

\begin{table*}[!t]
\centering
\caption{Yager's rule (IPEK) versus Dempster's rule (IPEK-Dempster) under dual-channel attacks. Open loop: mean AUC [lowest AUC over seeds; runs with AUC $< 0.5$]. Closed loop: mean F1, FPR and detection delay with frozen thresholds (0.39 and $10^{-4}$). $^{*}$: significantly different from Yager (paired Wilcoxon, $p < 0.05$); --: no attacker revoked.}
\label{tab:rule}
\scriptsize
\setlength{\tabcolsep}{2.4pt}
\begin{tabular}{llcccccccc}
\toprule
& & \multicolumn{2}{c}{\textbf{Open-loop AUC}} & \multicolumn{2}{c}{\textbf{F1}} & \multicolumn{2}{c}{\textbf{FPR (\%)}} & \multicolumn{2}{c}{\textbf{Delay (s)}} \\
\cmidrule(lr){3-4}\cmidrule(lr){5-6}\cmidrule(lr){7-8}\cmidrule(lr){9-10}
\textbf{Attack} & \textbf{$\pi_a$ (\%)} & \textbf{Yager} & \textbf{Dempster} & \textbf{Yager} & \textbf{Dempster} & \textbf{Yager} & \textbf{Dempster} & \textbf{Yager} & \textbf{Dempster} \\
\midrule
D-Sev & 10 & 0.976 [0.92; 0] & 0.983 [0.94; 0] & 0.897 & 0.951$^{*}$ & 1.9 & 0.1 & 187 & 217$^{*}$ \\
D-Sev & 25 & 0.904 [0.81; 0] & 0.961$^{*}$ [0.87; 0] & 0.845 & 0.911$^{*}$ & 7.8 & 1.0 & 251 & 289$^{*}$ \\
D-Sev & 30 & 0.801 [0.58; 0] & 0.889$^{*}$ [0.68; 0] & 0.735 & 0.791 & 16.9 & 6.4 & 290 & 329 \\
D-Sev & 35 & 0.560 [0.16; 5] & 0.601 [0.08; 4] & 0.533 & 0.484 & 41.8 & 29.3 & 273 & 322$^{*}$ \\
D-Sev & 40 & 0.366 [0.01; 7] & 0.333 [0.00; 7] & 0.300 & 0.258 & 67.9 & 55.8 & 247 & 345$^{*}$ \\
D-Loc & 10 & 0.993 [0.95; 0] & 0.994 [0.95; 0] & 0.929 & 0.981$^{*}$ & 1.6 & 0.1 & 196 & 224$^{*}$ \\
D-Loc & 25 & 0.959 [0.91; 0] & 0.988$^{*}$ [0.97; 0] & 0.871 & 0.956$^{*}$ & 8.4 & 0.6 & 239 & 282$^{*}$ \\
D-Loc & 30 & 0.880 [0.71; 0] & 0.947$^{*}$ [0.70; 0] & 0.836 & 0.888 & 13.7 & 3.5 & 272 & 327$^{*}$ \\
D-Loc & 35 & 0.614 [0.24; 4] & 0.584 [0.03; 4] & 0.566 & 0.542 & 50.6 & 42.3 & 272 & 323$^{*}$ \\
D-Loc & 40 & 0.375 [0.01; 6] & 0.266 [0.00; 7] & 0.368 & 0.220$^{*}$ & 71.2 & 73.3 & 249 & 337 \\
D-OnOff & 10 & 0.997 [0.99; 0] & 1.000$^{*}$ [1.00; 0] & 0.922 & 0.991$^{*}$ & 2.2 & 0.2 & 334 & 336 \\
D-OnOff & 25 & 0.876 [0.76; 0] & 0.781 [0.00; 2] & 0.764 & 0.715 & 15.5 & 18.5 & 445 & 522 \\
D-OnOff & 30 & 0.694 [0.50; 1] & 0.391 [0.00; 6] & 0.446 & 0.353 & 49.5 & 60.8 & 422 & 567 \\
D-OnOff & 35 & 0.415 [0.08; 6] & 0.065$^{*}$ [0.00; 10] & 0.154 & 0.028$^{*}$ & 85.0 & 93.4 & 371 & 541 \\
D-OnOff & 40 & 0.205 [0.00; 9] & 0.003$^{*}$ [0.00; 10] & 0.038 & 0.003$^{*}$ & 99.6 & 99.9 & 285 & 258 \\
D-Cont & 10 & 1.000 [1.00; 0] & 1.000 [1.00; 0] & 0.942 & 0.997$^{*}$ & 1.6 & 0.1 & 226 & 259$^{*}$ \\
D-Cont & 25 & 0.995 [0.98; 0] & 0.999 [0.99; 0] & 0.836 & 0.962$^{*}$ & 14.4 & 3.3 & 268 & 398$^{*}$ \\
D-Cont & 30 & 0.935 [0.82; 0] & 0.629 [0.00; 4] & 0.558 & 0.512 & 54.7 & 46.7 & 304 & 448$^{*}$ \\
D-Cont & 35 & 0.535 [0.03; 5] & 0.178$^{*}$ [0.00; 8] & 0.213 & 0.118 & 85.7 & 89.0 & 233 & 407 \\
D-Cont & 40 & 0.295 [0.00; 6] & 0.000$^{*}$ [0.00; 10] & 0.035 & 0.000$^{*}$ & 100.0 & 100.0 & 171 & -- \\
\bottomrule
\end{tabular}
\end{table*}

The comparison reveals a trade-off between peak performance and robustness. When honest evidence dominates, Dempster's rule performs better: under the severity- and location-triggered attacks up to 30\% it achieves a higher open-loop AUC (significantly at 25\% and 30\%), and with $\pi_a \leq 0.25$ it achieves a significantly higher closed-loop F1 in seven of the eight dual-channel configurations, with an FPR of at most 3.3\% except under D-OnOff at 25\%. In the reference configuration, for instance, IPEK-Dempster reaches an F1 of 0.911 at an FPR of 1.0\%, against 0.845 at 7.8\% for IPEK. Dempster's rule is, however, brittle. Its normalization sharpens whichever hypothesis leads (Proposition~\ref{prop:dogma}); with $\pi_a \leq 0.25$, the open-loop scores of the majority of attackers saturate below $10^{-12}$ in 15 of the 16 configurations, and once attackers prevail, the inversion is complete. Under the on-off and continuous attacks, its open-loop separation collapses from 25\%--30\% onward: over the 120 dual-channel runs with $\pi_a \leq 0.30$, 12 runs of IPEK-Dempster are inverted (AUC $< 0.5$), against a single run of IPEK. In the closed loop, IPEK-Dempster produces two runs with an FPR above 50\% at 25\% (both under D-OnOff, with F1 values of 0.00 and 0.18), whereas IPEK produces none. Yager's rule also revokes earlier: with $\pi_a \leq 0.25$, the delay of IPEK is significantly shorter in six of the eight dual-channel configurations, by 27--130\,s.

Yager's rule is therefore retained in IPEK for its worst-case behavior and its faster reaction rather than for its average performance. Where the attacker share is known to be low and the attack to follow criticality, Dempster's rule is a viable alternative that trades robustness for a lower FPR.

\subsection{Sensitivity and Strength of the Feedback Attack}\label{subsec:sens}
Table~\ref{tab:sens} varies each parameter of IPEK in the reference configuration, one at a time; the closed-loop column keeps the frozen threshold of 0.39.

\begin{table*}[!t]
\centering
\caption{Sensitivity of IPEK in the reference configuration (D-Sev, $\pi_a = 0.25$). Open loop: AUC and difference from the baseline. Closed loop (threshold 0.39): F1, difference from the baseline, recall, FPR and detection delay. $^{*}$: significant difference (paired Wilcoxon, $p < 0.05$). $N_r$ has no effect in the open loop.}
\label{tab:sens}
\footnotesize
\setlength{\tabcolsep}{3pt}
\begin{tabular}{lcccccccc}
\toprule
& & \multicolumn{2}{c}{\textbf{Open loop}} & \multicolumn{5}{c}{\textbf{Closed loop}} \\
\cmidrule(lr){3-4}\cmidrule(lr){5-9}
\textbf{Parameter} & \textbf{Value} & \textbf{AUC} & \textbf{$\Delta$} & \textbf{F1} & \textbf{$\Delta$} & \textbf{Recall} & \textbf{FPR (\%)} & \textbf{Delay (s)} \\
\midrule
Baseline & -- & 0.904 & -- & 0.845 & -- & 0.895 & 7.8 & 251 \\
\midrule
$\eta$ (base 5) & 2.5 & 0.832 & $-$0.072$^{*}$ & 0.596 & $-$0.249$^{*}$ & 0.475 & 2.5 & 337 \\
$\eta$ (base 5) & 7 & 0.904 & 0.000 & 0.701 & $-$0.143$^{*}$ & 0.950 & 26.1 & 222 \\
$\mu$ (base 0.15) & 0.075 & 0.863 & $-$0.041$^{*}$ & 0.796 & $-$0.049$^{*}$ & 0.835 & 9.0 & 289 \\
$\mu$ (base 0.15) & 0.20 & 0.904 & 0.000 & 0.823 & $-$0.021$^{*}$ & 0.889 & 9.3 & 252 \\
$N_r$ (base 3) & 1 & -- & -- & 0.757 & $-$0.088$^{*}$ & 0.982 & 21.4 & 137 \\
$N_r$ (base 3) & 5 & -- & -- & 0.785 & $-$0.060 & 0.698 & 2.4 & 368 \\
$\delta_0$ (base 0.5) & 0.3 & 0.944 & +0.040$^{*}$ & 0.862 & +0.017 & 0.886 & 5.8 & 247 \\
$\delta_0$ (base 0.5) & 0.7 & 0.818 & $-$0.086$^{*}$ & 0.726 & $-$0.119$^{*}$ & 0.829 & 15.8 & 281 \\
$\gamma$ (base 0.5) & 0.25 & 0.838 & $-$0.066$^{*}$ & 0.701 & $-$0.144$^{*}$ & 0.832 & 19.2 & 290 \\
$\gamma$ (base 0.5) & 1.0 & 0.932 & +0.028$^{*}$ & 0.902 & +0.057$^{*}$ & 0.908 & 3.7 & 228 \\
\bottomrule
\end{tabular}
\end{table*}

For the local-trust parameters $\eta$ and $\mu$, values above the baseline leave the separation unchanged (AUC 0.904 for $\eta = 7$ and $\mu = 0.20$), whereas smaller values reduce it (0.832 for $\eta = 2.5$ and 0.863 for $\mu = 0.075$): a sufficiently strong asymmetry is required, and beyond it the separation saturates. Part of the closed-loop variation is calibration rather than separation; with $\eta = 7$, for example, the AUC is unchanged but the FPR rises to 26.1\% at the frozen threshold, since the scores shift downward. The number of consecutive rounds $N_r$ adjusts the balance between recall, FPR and delay: with $N_r = 1$, the delay drops to 137\,s and the FPR rises to 21.4\%; with $N_r = 5$, the FPR drops to 2.4\% and the delay grows to 368\,s.

The credibility parameters $\gamma$ and $\delta_0$ were fixed before the evaluation. Stricter settings would have performed better in the reference configuration: $\gamma = 1$ raises the closed-loop F1 to 0.902 and $\delta_0 = 0.3$ raises the AUC to 0.944. The reported results therefore do not stem from a tuned optimum; the effect of these settings on the other attacks was not evaluated, and the values were not changed after the analysis.

\begin{figure}[!t]
\centering
\includegraphics[width=\columnwidth]{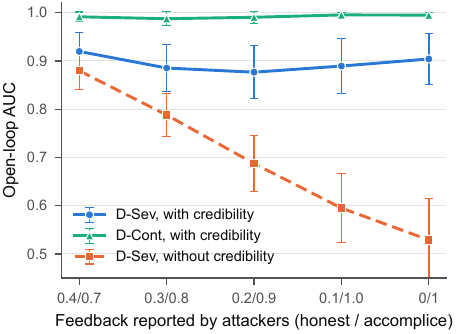}
\caption{Open-loop AUC of IPEK under feedback attacks of increasing strength ($\pi_a = 0.25$; mean and 95\% confidence interval over ten seeds). The horizontal axis gives the values reported by attackers for honest vehicles and for accomplices.}
\label{fig:severity}
\end{figure}

Fig.~\ref{fig:severity} varies the strength of the feedback manipulation under D-Sev and D-Cont at 25\%, from moderate values (0.4 for honest vehicles and 0.7 for accomplices) to the extreme values used elsewhere in the evaluation.

Without reporter credibility, the separation degrades almost linearly with the strength of the manipulation, from 0.879 to 0.529. With reporter credibility, it does not fall below 0.877 at any strength. The shallow minimum at intermediate strength reflects a natural trade-off for the attacker: moderate values deviate less from the consensus and partly escape the deviation test (the mean credibility of attackers is 0.53 at 0.2/0.9, compared with 0.42 at 0.1/1.0 and 0.75 for honest vehicles), but they also inflict less damage.

\subsection{Communication Overhead}\label{subsec:overhead}
In the open loop, the numbers of event reports and feedback messages are identical for the three methods in every run, with an average of 24\,482 reports and 23\,804 feedback messages per run (150 vehicles, 1000\,s). IPEK introduces no additional messages; the only difference lies in the payload of the feedback, which carries continuous local trust values instead of binary ones. In the closed loop, the message counts depend on the number of revoked vehicles and are therefore not a property of the method. The additional computation at the CA, one median per target and round for the credibility test, is analyzed in Section~\ref{subsec:cost}.
\section{Discussion and Limitations}\label{sec:discussion}

\textbf{Why IPEK works against strategic attackers.} The results support the design rationale of Section~\ref{sec:ipek}. The strategic attacker builds reputation on trivial events and spends it on critical ones; IPEK makes this exchange unprofitable, since a single lie on a critical event costs more trust than dozens of correct reports on trivial events can restore (Proposition~\ref{prop:comp}). The attacker that also corrupts its feedback is neutralized by the credibility test, which detects the pattern of its feedback rather than its values and thereby renders the strength of the manipulation largely irrelevant (Fig.~\ref{fig:severity}). The evidential fusion with Yager's rule keeps the global state revisable, so that conflicting evidence does not lock a vehicle into an irreversible verdict (Proposition~\ref{prop:dogma}). Together, these mechanisms yield an FPR that remains low across all attacks with up to 25\% attackers, which is the property most relevant to deployment, since every false revocation removes an honest participant from the network.

\textbf{Operating envelope.} Under fully colluding dual-channel attacks, IPEK is reliable up to a realized attacker share of about one third and fails beyond it by inversion. This limit is not specific to IPEK: every method that infers trust from the consensus of witnesses loses its reference when attackers become the majority among them, and the baselines fail in the same region. Under report-only attacks, no such limit was reached up to 40\%. Extending the envelope requires evidence that does not depend on the witnesses, such as direct observations by the infrastructure, which is a planned direction of future work. The full-collusion assumption is deliberately pessimistic; attackers that do not coordinate would not form a consistent majority as early.

\textbf{On-off attacks.} The on-off attack, which ignores event importance, is the weakest case of IPEK. With up to 25\% attackers, IPEK still outperforms both baselines in the closed loop and separates at least as well as beta reputation in the open loop; at 30\% and above, beta reputation degrades more gracefully. The weakness is already visible in the open loop, where no revocation takes place, and therefore originates in the trust computation rather than in the revocation rule; and part of it is the measured cost of criticality awareness when the attack is unrelated to criticality (Section~\ref{subsec:ablation}). A detailed diagnosis is left for future work.

\textbf{Detection delay.} IPEK revokes attackers later than beta reputation. This is the price of requiring consistent evidence over several rounds and of the cautious handling of conflict, and it is accompanied by a substantially lower FPR. The delay is controlled by $N_r$, which offers a direct trade-off between reaction time and false revocations (Table~\ref{tab:sens}).

\textbf{Threshold transfer.} All closed-loop results use a single threshold per method, selected in one configuration. The distribution of honest trust, however, shifts with the attack and the attacker share; under the continuous dual-channel attack at 25\%, for example, IPEK separates attackers and honest vehicles almost perfectly in the open loop, yet its frozen threshold produces an FPR of 14.4\%. The same phenomenon affects the baselines more strongly, which contributes to the closed-loop margins of IPEK under continuous attacks (Section~\ref{subsec:closed}). A threshold that adapts to the observed trust distribution is a natural extension.

\textbf{Choice of combination rule.} Dempster's rule yields higher peak performance when honest evidence dominates and the attack follows criticality, whereas Yager's rule avoids the irreversible inversions that Dempster's rule produces once attackers prevail. IPEK adopts Yager's rule for its worst-case behavior; in deployments where the attacker share is known to be low, Dempster's rule may be preferred.

\subsection{Threats to Validity}\label{subsec:threats}
Several limitations of the experimental design should be considered when interpreting these findings. Most of them, however, concern the \emph{external validity} (generalizability) of the results rather than the \emph{internal validity} of the comparison, since all shared modeling choices affect the compared methods identically, keeping the comparison fair.

The vehicle density in the simulation is low relative to the network area (150 vehicles on a 2000\,$\times$\,2000\,m grid), producing a largely free-flow traffic profile. Free-flow conditions shorten contact times and thereby reduce the frequency of repeated interactions between vehicles. Since repeated interaction is a natural component of trust accumulation, this creates a conservative operating environment that makes all trust mechanisms work harder. As this effect is identical for all methods, it does not compromise the validity of the observed relative performance differences. Nonetheless, absolute performance values---particularly detection rates---should not be directly generalized to dense, congested urban traffic profiles.

As noted in Section~\ref{subsec:simenv}, a single RSU is modeled as a central certificate authority, and coverage limits and multi-RSU coordination are not addressed. This abstraction does not affect the comparison, since all compared methods are centralized; however, the findings should be re-evaluated along these dimensions before being transferred to distributed or coverage-limited deployments.

Messages are delivered without a physical- or medium-access-layer model. Packet losses and delays would affect all methods, but they could affect them differently, for instance by delaying the feedback on which revocation depends; the results make no claim about performance under realistic channel conditions. Likewise, a single topology, a single traffic density and a single event schedule were used.

The design of IPEK and its evaluation share the same scenario family. In particular, the reporter-credibility component was introduced after its need had been observed under the dual-channel attacks of this scenario. Its parameters were fixed in advance and not tuned, and the sensitivity analysis indicates that the reported configuration is not a tuned optimum, but an independent validation scenario was not used.

Each configuration was run with ten seeds, and the realized attacker share varies between seeds; this variation is reported explicitly (Section~\ref{subsec:break}). Statistical tests were not corrected for multiple comparisons in general. The main claims, i.e., the superiority of IPEK with up to 25\% attackers, remain significant after the Holm correction, whereas individual results with $p$ values close to 0.05 should be read as indicative. Finally, attackers that remain silent cannot be observed by any report-based trust mechanism and are outside the scope of this study.
\section{Conclusion}\label{sec:conclusion}
This study presented IPEK, a trust management framework targeting a vulnerability of event-based trust in VANETs: insider attackers who behave strategically according to event criticality, building reputation on trivial events and exploiting it on critical ones, while corrupting the trust feedback they provide about others. IPEK addresses this threat with four components: an asymmetric local trust that scales rewards and penalties with event severity and location criticality, a reporter-credibility test that discounts feedback deviating systematically from the consensus, an evidential global trust that fuses feedback with Yager's rule, historical combination and risk accentuation, and a revocation rule that requires consistently low trust over several rounds. The analytical properties of the update rules quantify the asymmetry that makes the strategic attack unprofitable and show why Yager's rule keeps the global state revisable.

The evaluation compared IPEK with TCEMD and beta reputation under eight attacks, combining four triggers with dual-channel and report-only falsification, and attacker ratios from 10\% to 40\%, using a threshold-free open-loop layer and a closed-loop layer with thresholds frozen in a reference configuration. With up to 25\% attackers, IPEK achieved a significantly higher F1 than both baselines in all 16 configurations, with an FPR of 0.9\%--15.5\% against 7.8\%--49.7\% for beta reputation and 26.0\%--41.4\% for TCEMD; against report-only attacks, it remained robust up to 40\% attackers. Reporter credibility proved indispensable against feedback manipulation and made the separation insensitive to the strength of the manipulation, and criticality awareness improved detection under the attacks that follow criticality at a small, measured cost under those that do not. The evaluation also delineated the limits of the approach: under fully colluding dual-channel attacks, all compared methods break down once attackers make up about one third of the vehicles, IPEK revokes attackers later than beta reputation, and the on-off attack is its weakest case at higher attacker ratios.

Future work will proceed in four directions: (i) anchoring trust in evidence that does not depend on the witnesses, such as infrastructure observations, to extend the operating envelope beyond one third of attackers; (ii) a detection threshold that adapts to the observed trust distribution; (iii) a diagnosis of and a remedy for the on-off weakness; and (iv) an evaluation under realistic channel conditions, multiple RSUs and denser traffic.

\backmatter

\bmhead{Acknowledgements}
The author gratefully acknowledges the Scientific and Technological Research Council of T\"{u}rkiye (T\"{U}B\.{I}TAK) for its support of this research.

\section*{Statements and Declarations}

\bmhead{Funding}
This work was supported by the Scientific and Technological Research Council of T\"{u}rkiye (T\"{U}B\.{I}TAK) under Grant Number 124E017.

\bmhead{Competing interests}
The author declares no competing financial or non-financial interests.

\bmhead{Author contributions}
\.{I}pek Abas{\i}kele\c{s}-Turgut is the sole author of this work and was responsible for the conception and design of the method, the implementation of the simulation code, the experiments, the analysis and interpretation of the results, and the writing of the manuscript. The author read and approved the final manuscript.

\bmhead{Data availability}
The outputs of all 4080 simulation runs analysed in this study, a per-run table of the reported metrics, and the scripts that compute the metrics and statistical tests from the raw outputs are openly available in Zenodo at https://doi.org/10.5281/zenodo.XXXXXXX.

\bmhead{Code availability}
The simulation code is not publicly available, since it forms part of an ongoing funded research project.

\bmhead{Ethics approval and consent to participate}
Not applicable.


\begin{thebibliography}{33}
\ifx \bisbn   \undefined \def \bisbn  #1{ISBN #1}\fi
\ifx \binits  \undefined \def \binits#1{#1}\fi
\ifx \bauthor  \undefined \def \bauthor#1{#1}\fi
\ifx \batitle  \undefined \def \batitle#1{#1}\fi
\ifx \bjtitle  \undefined \def \bjtitle#1{#1}\fi
\ifx \bvolume  \undefined \def \bvolume#1{\textbf{#1}}\fi
\ifx \byear  \undefined \def \byear#1{#1}\fi
\ifx \bissue  \undefined \def \bissue#1{#1}\fi
\ifx \bfpage  \undefined \def \bfpage#1{#1}\fi
\ifx \blpage  \undefined \def \blpage #1{#1}\fi
\ifx \burl  \undefined \def \burl#1{\textsf{#1}}\fi
\ifx \doiurl  \undefined \def \doiurl#1{\url{https://doi.org/#1}}\fi
\ifx \betal  \undefined \def \betal{\textit{et al.}}\fi
\ifx \binstitute  \undefined \def \binstitute#1{#1}\fi
\ifx \binstitutionaled  \undefined \def \binstitutionaled#1{#1}\fi
\ifx \bctitle  \undefined \def \bctitle#1{#1}\fi
\ifx \beditor  \undefined \def \beditor#1{#1}\fi
\ifx \bpublisher  \undefined \def \bpublisher#1{#1}\fi
\ifx \bbtitle  \undefined \def \bbtitle#1{#1}\fi
\ifx \bedition  \undefined \def \bedition#1{#1}\fi
\ifx \bseriesno  \undefined \def \bseriesno#1{#1}\fi
\ifx \blocation  \undefined \def \blocation#1{#1}\fi
\ifx \bsertitle  \undefined \def \bsertitle#1{#1}\fi
\ifx \bsnm \undefined \def \bsnm#1{#1}\fi
\ifx \bsuffix \undefined \def \bsuffix#1{#1}\fi
\ifx \bparticle \undefined \def \bparticle#1{#1}\fi
\ifx \barticle \undefined \def \barticle#1{#1}\fi
\bibcommenthead
\ifx \bconfdate \undefined \def \bconfdate #1{#1}\fi
\ifx \botherref \undefined \def \botherref #1{#1}\fi
\ifx \url \undefined \def \url#1{\textsf{#1}}\fi
\ifx \bchapter \undefined \def \bchapter#1{#1}\fi
\ifx \bbook \undefined \def \bbook#1{#1}\fi
\ifx \bcomment \undefined \def \bcomment#1{#1}\fi
\ifx \oauthor \undefined \def \oauthor#1{#1}\fi
\ifx \citeauthoryear \undefined \def \citeauthoryear#1{#1}\fi
\ifx \endbibitem  \undefined \def \endbibitem {}\fi
\ifx \bconflocation  \undefined \def \bconflocation#1{#1}\fi
\ifx \arxivurl  \undefined \def \arxivurl#1{\textsf{#1}}\fi
\csname PreBibitemsHook\endcsname

\bibitem[\protect\citeauthoryear{Abas{\i}kele{\c{s}}-Turgut}{2024}]{abasikelecs2024recent}
\begin{bchapter}
\bauthor{\bsnm{Abas{\i}kele{\c{s}}-Turgut}, \binits{{\. I}.}}:
\bctitle{Recent reputation-based security approaches for {VANET}s}.
In: \bbtitle{2024 8th International Artificial Intelligence and Data Processing
  Symposium (IDAP)},
pp. \bfpage{1}--\blpage{6}
(\byear{2024}).
\doiurl{10.1109/IDAP64064.2024.10710994}
\end{bchapter}
\endbibitem

\bibitem[\protect\citeauthoryear{{Business of Apps}}{2026}]{businessofapps2026}
\begin{botherref}
\oauthor{\bsnm{{Business of Apps}}}:
Navigation App Revenue and Usage Statistics (2026).
Accessed 4 March 2026
(2026).
\url{https://www.businessofapps.com/data/navigation-app-market/}
\end{botherref}
\endbibitem

\bibitem[\protect\citeauthoryear{Pichai}{2024}]{pichai2024alphabet}
\begin{botherref}
\oauthor{\bsnm{Pichai}, \binits{S.}}:
Alphabet {Q3} 2024 Earnings Call.
Alphabet Inc.
(2024).
\url{https://blog.google/inside-google/message-ceo/alphabet-earnings-q3-2024/}
\end{botherref}
\endbibitem

\bibitem[\protect\citeauthoryear{Bradshaw}{2019}]{bradshaw2019googlemaps}
\begin{botherref}
\oauthor{\bsnm{Bradshaw}, \binits{K.}}:
{G}oogle {M}aps Expands {W}aze-like Crowdsourcing with New Incident Types.
Android Police
(2019).
\url{https://www.androidpolice.com/2019/10/20/google-maps-adds-waze-like-crowdsourcing-with-4-new-incident-types/}
\end{botherref}
\endbibitem

\bibitem[\protect\citeauthoryear{Liu et~al.}{2020}]{tcemd}
\begin{barticle}
\bauthor{\bsnm{Liu}, \binits{Z.}},
\bauthor{\bsnm{Weng}, \binits{J.}},
\bauthor{\bsnm{Ma}, \binits{J.}},
\bauthor{\bsnm{Guo}, \binits{J.}},
\bauthor{\bsnm{Feng}, \binits{B.}},
\bauthor{\bsnm{Jiang}, \binits{Z.}},
\bauthor{\bsnm{Wei}, \binits{K.}}:
\batitle{{TCEMD}: A trust cascading-based emergency message dissemination model
  in {VANET}s}.
\bjtitle{IEEE Internet of Things Journal}
\bvolume{7}(\bissue{5}),
\bfpage{4028}--\blpage{4048}
(\byear{2020})
\doiurl{10.1109/JIOT.2019.2957520}
\end{barticle}
\endbibitem

\bibitem[\protect\citeauthoryear{Ahmad et~al.}{2021}]{notrino}
\begin{barticle}
\bauthor{\bsnm{Ahmad}, \binits{F.}},
\bauthor{\bsnm{Kurugollu}, \binits{F.}},
\bauthor{\bsnm{Kerrache}, \binits{C.A.}},
\bauthor{\bsnm{Sezer}, \binits{S.}},
\bauthor{\bsnm{Liu}, \binits{L.}}:
\batitle{{NOTRINO}: A novel hybrid trust management scheme for
  internet-of-vehicles}.
\bjtitle{IEEE Transactions on Vehicular Technology}
\bvolume{70}(\bissue{9}),
\bfpage{9244}--\blpage{9257}
(\byear{2021})
\doiurl{10.1109/TVT.2021.3049189}
\end{barticle}
\endbibitem

\bibitem[\protect\citeauthoryear{Bhargava and Verma}{2022}]{duel}
\begin{barticle}
\bauthor{\bsnm{Bhargava}, \binits{A.}},
\bauthor{\bsnm{Verma}, \binits{S.}}:
\batitle{{DUEL}: {D}empster uncertainty-based enhanced-trust level scheme for
  {VANET}}.
\bjtitle{IEEE Transactions on Intelligent Transportation Systems}
\bvolume{23}(\bissue{9}),
\bfpage{15079}--\blpage{15090}
(\byear{2022})
\doiurl{10.1109/TITS.2021.3136548}
\end{barticle}
\endbibitem

\bibitem[\protect\citeauthoryear{Qi et~al.}{2024}]{htemd}
\begin{barticle}
\bauthor{\bsnm{Qi}, \binits{J.}},
\bauthor{\bsnm{Zheng}, \binits{N.}},
\bauthor{\bsnm{Xu}, \binits{M.}},
\bauthor{\bsnm{Chen}, \binits{P.}},
\bauthor{\bsnm{Li}, \binits{W.}}:
\batitle{A hybrid-trust-based emergency message dissemination model for
  vehicular ad hoc networks}.
\bjtitle{Journal of Information Security and Applications}
\bvolume{81},
\bfpage{103699}
(\byear{2024})
\doiurl{10.1016/j.jisa.2024.103699}
\end{barticle}
\endbibitem

\bibitem[\protect\citeauthoryear{Atwa et~al.}{2021}]{rteam}
\begin{barticle}
\bauthor{\bsnm{Atwa}, \binits{R.J.}},
\bauthor{\bsnm{Flocchini}, \binits{P.}},
\bauthor{\bsnm{Nayak}, \binits{A.}}:
\batitle{{RTEAM}: Risk-based trust evaluation advanced model for {VANET}s}.
\bjtitle{IEEE Access}
\bvolume{9},
\bfpage{117772}--\blpage{117783}
(\byear{2021})
\doiurl{10.1109/ACCESS.2021.3107467}
\end{barticle}
\endbibitem

\bibitem[\protect\citeauthoryear{Qi et~al.}{2023}]{mdt}
\begin{barticle}
\bauthor{\bsnm{Qi}, \binits{J.}},
\bauthor{\bsnm{Zheng}, \binits{N.}},
\bauthor{\bsnm{Xu}, \binits{M.}},
\bauthor{\bsnm{Wang}, \binits{X.}},
\bauthor{\bsnm{Chen}, \binits{Y.}}:
\batitle{A multi-dimensional trust model for misbehavior detection in vehicular
  ad hoc networks}.
\bjtitle{Journal of Information Security and Applications}
\bvolume{76},
\bfpage{103528}
(\byear{2023})
\doiurl{10.1016/j.jisa.2023.103528}
\end{barticle}
\endbibitem

\bibitem[\protect\citeauthoryear{Liu et~al.}{2022}]{hdrs}
\begin{barticle}
\bauthor{\bsnm{Liu}, \binits{X.}},
\bauthor{\bsnm{Ma}, \binits{O.}},
\bauthor{\bsnm{Chen}, \binits{W.}},
\bauthor{\bsnm{Xia}, \binits{Y.}},
\bauthor{\bsnm{Zhou}, \binits{Y.}}:
\batitle{{HDRS}: A hybrid reputation system with dynamic update interval for
  detecting malicious vehicles in {VANET}s}.
\bjtitle{IEEE Transactions on Intelligent Transportation Systems}
\bvolume{23}(\bissue{8}),
\bfpage{12766}--\blpage{12777}
(\byear{2022})
\doiurl{10.1109/TITS.2021.3117289}
\end{barticle}
\endbibitem

\bibitem[\protect\citeauthoryear{Ahmad et~al.}{2020}]{marine}
\begin{barticle}
\bauthor{\bsnm{Ahmad}, \binits{F.}},
\bauthor{\bsnm{Kurugollu}, \binits{F.}},
\bauthor{\bsnm{Adnane}, \binits{A.}},
\bauthor{\bsnm{Hussain}, \binits{R.}},
\bauthor{\bsnm{Hussain}, \binits{F.}}:
\batitle{{MARINE}: Man-in-the-middle attack resistant trust model in connected
  vehicles}.
\bjtitle{IEEE Internet of Things Journal}
\bvolume{7}(\bissue{4}),
\bfpage{3310}--\blpage{3322}
(\byear{2020})
\doiurl{10.1109/JIOT.2020.2967568}
\end{barticle}
\endbibitem

\bibitem[\protect\citeauthoryear{Zhang et~al.}{2020}]{aatms}
\begin{barticle}
\bauthor{\bsnm{Zhang}, \binits{J.}},
\bauthor{\bsnm{Zheng}, \binits{K.}},
\bauthor{\bsnm{Zhang}, \binits{D.}},
\bauthor{\bsnm{Yan}, \binits{B.}}:
\batitle{{AATMS}: An anti-attack trust management scheme in {VANET}}.
\bjtitle{IEEE Access}
\bvolume{8},
\bfpage{21077}--\blpage{21090}
(\byear{2020})
\doiurl{10.1109/ACCESS.2020.2966747}
\end{barticle}
\endbibitem

\bibitem[\protect\citeauthoryear{Cui et~al.}{2019}]{rsma}
\begin{barticle}
\bauthor{\bsnm{Cui}, \binits{J.}},
\bauthor{\bsnm{Zhang}, \binits{X.}},
\bauthor{\bsnm{Zhong}, \binits{H.}},
\bauthor{\bsnm{Ying}, \binits{Z.}},
\bauthor{\bsnm{Liu}, \binits{L.}}:
\batitle{{RSMA}: Reputation system-based lightweight message authentication
  framework and protocol for {5G}-enabled vehicular networks}.
\bjtitle{IEEE Internet of Things Journal}
\bvolume{6}(\bissue{4}),
\bfpage{6417}--\blpage{6428}
(\byear{2019})
\doiurl{10.1109/JIOT.2019.2895136}
\end{barticle}
\endbibitem

\bibitem[\protect\citeauthoryear{El~Sayed et~al.}{2020}]{htms}
\begin{barticle}
\bauthor{\bsnm{El~Sayed}, \binits{H.}},
\bauthor{\bsnm{Zeadally}, \binits{S.}},
\bauthor{\bsnm{Puthal}, \binits{D.}}:
\batitle{Design and evaluation of a novel hierarchical trust assessment
  approach for vehicular networks}.
\bjtitle{Vehicular Communications}
\bvolume{24},
\bfpage{100227}
(\byear{2020})
\doiurl{10.1016/j.vehcom.2019.100227}
\end{barticle}
\endbibitem

\bibitem[\protect\citeauthoryear{Zhang et~al.}{2018}]{misbehav}
\begin{barticle}
\bauthor{\bsnm{Zhang}, \binits{C.}},
\bauthor{\bsnm{Chen}, \binits{K.}},
\bauthor{\bsnm{Zeng}, \binits{X.}},
\bauthor{\bsnm{Xue}, \binits{X.}}:
\batitle{Misbehavior detection based on support vector machine and
  {D}empster--{S}hafer theory of evidence in {VANET}s}.
\bjtitle{IEEE Access}
\bvolume{6},
\bfpage{59860}--\blpage{59870}
(\byear{2018})
\doiurl{10.1109/ACCESS.2018.2875678}
\end{barticle}
\endbibitem

\bibitem[\protect\citeauthoryear{Turgut}{2025}]{turgut2025effect}
\begin{bchapter}
\bauthor{\bsnm{Turgut}, \binits{{\. I}.A.}}:
\bctitle{The effect of selfish nodes on the performance of event-based trust
  management in {VANET}s}.
In: \bbtitle{2025 9th International Symposium on Innovative Approaches in Smart
  Technologies (ISAS)},
pp. \bfpage{1}--\blpage{5}
(\byear{2025}).
\doiurl{10.1109/ISAS66241.2025.11101836}
\end{bchapter}
\endbibitem

\bibitem[\protect\citeauthoryear{Dempster}{1967}]{dempster1967}
\begin{barticle}
\bauthor{\bsnm{Dempster}, \binits{A.P.}}:
\batitle{Upper and lower probabilities induced by a multivalued mapping}.
\bjtitle{The Annals of Mathematical Statistics}
\bvolume{38}(\bissue{2}),
\bfpage{325}--\blpage{339}
(\byear{1967})
\doiurl{10.1214/aoms/1177698950}
\end{barticle}
\endbibitem

\bibitem[\protect\citeauthoryear{Shafer}{1976}]{shafer1976}
\begin{bbook}
\bauthor{\bsnm{Shafer}, \binits{G.}}:
\bbtitle{A Mathematical Theory of Evidence}.
\bpublisher{Princeton University Press},
\blocation{Princeton, NJ}
(\byear{1976}).
\doiurl{10.1515/9780691214696}
\end{bbook}
\endbibitem

\bibitem[\protect\citeauthoryear{Zadeh}{1986}]{zadeh1986}
\begin{barticle}
\bauthor{\bsnm{Zadeh}, \binits{L.A.}}:
\batitle{A simple view of the {D}empster--{S}hafer theory of evidence and its
  implication for the rule of combination}.
\bjtitle{AI Magazine}
\bvolume{7}(\bissue{2}),
\bfpage{85}--\blpage{90}
(\byear{1986})
\doiurl{10.1609/aimag.v7i2.542}
\end{barticle}
\endbibitem

\bibitem[\protect\citeauthoryear{Yager}{1987}]{yager1987}
\begin{barticle}
\bauthor{\bsnm{Yager}, \binits{R.R.}}:
\batitle{On the {D}empster--{S}hafer framework and new combination rules}.
\bjtitle{Information Sciences}
\bvolume{41}(\bissue{2}),
\bfpage{93}--\blpage{137}
(\byear{1987})
\doiurl{10.1016/0020-0255(87)90007-7}
\end{barticle}
\endbibitem

\bibitem[\protect\citeauthoryear{Smarandache and
  Dezert}{2005}]{smarandache2005pcr}
\begin{bchapter}
\bauthor{\bsnm{Smarandache}, \binits{F.}},
\bauthor{\bsnm{Dezert}, \binits{J.}}:
\bctitle{Information fusion based on new proportional conflict redistribution
  rules}.
In: \bbtitle{2005 7th International Conference on Information Fusion (FUSION)}
(\byear{2005}).
\doiurl{10.1109/ICIF.2005.1591955}
\end{bchapter}
\endbibitem

\bibitem[\protect\citeauthoryear{Raya et~al.}{2008}]{raya2008datacentric}
\begin{bchapter}
\bauthor{\bsnm{Raya}, \binits{M.}},
\bauthor{\bsnm{Papadimitratos}, \binits{P.}},
\bauthor{\bsnm{Gligor}, \binits{V.D.}},
\bauthor{\bsnm{Hubaux}, \binits{J.-P.}}:
\bctitle{On data-centric trust establishment in ephemeral ad hoc networks}.
In: \bbtitle{IEEE INFOCOM 2008 -- The 27th Conference on Computer
  Communications},
pp. \bfpage{1238}--\blpage{1246}
(\byear{2008}).
\doiurl{10.1109/INFOCOM.2008.180}
\end{bchapter}
\endbibitem

\bibitem[\protect\citeauthoryear{J{\o}sang and Ismail}{2002}]{josang2002beta}
\begin{bchapter}
\bauthor{\bsnm{J{\o}sang}, \binits{A.}},
\bauthor{\bsnm{Ismail}, \binits{R.}}:
\bctitle{The beta reputation system}.
In: \bbtitle{Proceedings of the 15th Bled Electronic Commerce Conference},
\bconflocation{Bled, Slovenia}
(\byear{2002})
\end{bchapter}
\endbibitem

\bibitem[\protect\citeauthoryear{J{\o}sang}{2016}]{josang2016sl}
\begin{bbook}
\bauthor{\bsnm{J{\o}sang}, \binits{A.}}:
\bbtitle{Subjective Logic: A Formalism for Reasoning Under Uncertainty}.
\bpublisher{Springer},
\blocation{Cham}
(\byear{2016}).
\doiurl{10.1007/978-3-319-42337-1}
\end{bbook}
\endbibitem

\bibitem[\protect\citeauthoryear{Whitby et~al.}{2004}]{whitby2004filtering}
\begin{bchapter}
\bauthor{\bsnm{Whitby}, \binits{A.}},
\bauthor{\bsnm{J{\o}sang}, \binits{A.}},
\bauthor{\bsnm{Indulska}, \binits{J.}}:
\bctitle{Filtering out unfair ratings in {B}ayesian reputation systems}.
In: \bbtitle{Proceedings of the 7th International Workshop on Trust in Agent
  Societies},
\bconflocation{New York, USA}
(\byear{2004})
\end{bchapter}
\endbibitem

\bibitem[\protect\citeauthoryear{Dellarocas}{2000}]{dellarocas2000immunizing}
\begin{bchapter}
\bauthor{\bsnm{Dellarocas}, \binits{C.}}:
\bctitle{Immunizing online reputation reporting systems against unfair ratings
  and discriminatory behavior}.
In: \bbtitle{Proceedings of the 2nd ACM Conference on Electronic Commerce (EC
  '00)},
pp. \bfpage{150}--\blpage{157}
(\byear{2000}).
\doiurl{10.1145/352871.352889}
\end{bchapter}
\endbibitem

\bibitem[\protect\citeauthoryear{Varga and Hornig}{2008}]{varga2008omnet}
\begin{bchapter}
\bauthor{\bsnm{Varga}, \binits{A.}},
\bauthor{\bsnm{Hornig}, \binits{R.}}:
\bctitle{An overview of the {OMNeT++} simulation environment}.
In: \bbtitle{Proceedings of the 1st International Conference on Simulation
  Tools and Techniques for Communications, Networks and Systems \& Workshops
  (SIMUTools)}
(\byear{2008}).
\doiurl{10.4108/ICST.SIMUTOOLS2008.3027}
\end{bchapter}
\endbibitem

\bibitem[\protect\citeauthoryear{Sommer et~al.}{2011}]{sommer2011veins}
\begin{barticle}
\bauthor{\bsnm{Sommer}, \binits{C.}},
\bauthor{\bsnm{German}, \binits{R.}},
\bauthor{\bsnm{Dressler}, \binits{F.}}:
\batitle{Bidirectionally coupled network and road traffic simulation for
  improved {IVC} analysis}.
\bjtitle{IEEE Transactions on Mobile Computing}
\bvolume{10}(\bissue{1}),
\bfpage{3}--\blpage{15}
(\byear{2011})
\doiurl{10.1109/TMC.2010.133}
\end{barticle}
\endbibitem

\bibitem[\protect\citeauthoryear{Lopez et~al.}{2018}]{lopez2018sumo}
\begin{bchapter}
\bauthor{\bsnm{Lopez}, \binits{P.A.}},
\bauthor{\bsnm{Behrisch}, \binits{M.}},
\bauthor{\bsnm{Bieker-Walz}, \binits{L.}},
\bauthor{\bsnm{Erdmann}, \binits{J.}},
\bauthor{\bsnm{Fl{\"o}tter{\"o}d}, \binits{Y.-P.}},
\bauthor{\bsnm{Hilbrich}, \binits{R.}},
\bauthor{\bsnm{L{\"u}cken}, \binits{L.}},
\bauthor{\bsnm{Rummel}, \binits{J.}},
\bauthor{\bsnm{Wagner}, \binits{P.}},
\bauthor{\bsnm{Wie{\ss}ner}, \binits{E.}}:
\bctitle{Microscopic traffic simulation using {SUMO}}.
In: \bbtitle{2018 21st International Conference on Intelligent Transportation
  Systems (ITSC)},
pp. \bfpage{2575}--\blpage{2582}
(\byear{2018}).
\doiurl{10.1109/ITSC.2018.8569938}
\end{bchapter}
\endbibitem

\bibitem[\protect\citeauthoryear{Fawcett}{2006}]{fawcett2006roc}
\begin{barticle}
\bauthor{\bsnm{Fawcett}, \binits{T.}}:
\batitle{An introduction to {ROC} analysis}.
\bjtitle{Pattern Recognition Letters}
\bvolume{27}(\bissue{8}),
\bfpage{861}--\blpage{874}
(\byear{2006})
\doiurl{10.1016/j.patrec.2005.10.010}
\end{barticle}
\endbibitem

\bibitem[\protect\citeauthoryear{Wilcoxon}{1945}]{wilcoxon1945}
\begin{barticle}
\bauthor{\bsnm{Wilcoxon}, \binits{F.}}:
\batitle{Individual comparisons by ranking methods}.
\bjtitle{Biometrics Bulletin}
\bvolume{1}(\bissue{6}),
\bfpage{80}--\blpage{83}
(\byear{1945})
\doiurl{10.2307/3001968}
\end{barticle}
\endbibitem

\bibitem[\protect\citeauthoryear{Holm}{1979}]{holm1979}
\begin{barticle}
\bauthor{\bsnm{Holm}, \binits{S.}}:
\batitle{A simple sequentially rejective multiple test procedure}.
\bjtitle{Scandinavian Journal of Statistics}
\bvolume{6}(\bissue{2}),
\bfpage{65}--\blpage{70}
(\byear{1979})
\end{barticle}
\endbibitem

\end{thebibliography}
\end{document}